\documentclass[11pt,a4paper]{article}
\pdfoutput=1
\usepackage{jheppub}
\usepackage{gensymb}
\usepackage{placeins}
\usepackage{subfigure}
\usepackage{amssymb,amsmath}
\usepackage{graphicx}
\usepackage{color}
\usepackage{cancel}
\usepackage[colorlinks=true
,urlcolor=blue
,citecolor=blue
,linkcolor=blue
,pagecolor=blue
,linktocpage=true
,pdfproducer=medialab
]{hyperref}
 \usepackage[numbers]{natbib}
\usepackage{notoccite}
\makeatletter \renewcommand{\@dotsep}{10000} \makeatother
\def\be{\begin{equation}}
\def\ee{\end{equation}}
\def\bea{\begin{eqnarray}}
\def\eea{\end{eqnarray}}
\def\bi{\begin{itemize}}
\def\ei{\end{itemize}}

\usepackage[nodisplayskipstretch]{setspace}

\def\mgut{M_{\rm GUT}}

\newcommand{\beq}{\begin{equation}}
\newcommand{\eeq}{\end{equation}}

\begin{document}

\begin{titlepage}
\pagestyle{empty}

\vspace*{0.2in}
\begin{center}
{\Large \bf   Sparticle Spectroscopy at LHC-Run3 and LSP Dark Matter in light of Muon $\mathbf{g-2}$
  }\\
\vspace{1cm}
{\bf  Qaisar Shafi$^{a,}$\footnote{E-mail: qshafi@udel.edu} {\rm and}
Cem Salih $\ddot{\rm U}$n$^{b,}\hspace{0.05cm}$\footnote{E-mail: cemsalihun@uludag.edu.tr}}
\vspace{0.5cm}

{\it
$^a$Department of Physics and Astronomy,
University of Delaware, Newark, DE 19716, USA \\
$^b$Department of Physics, Bursa Uluda\~{g} University, TR16059 Bursa, Turkey
}

\end{center}

\vspace{0.5cm}
\begin{abstract}

Inspired by the latest measurement of muon g-2 by the Fermilab Experiment, we revisit a class of supersymmetric models in which non-universality at $\mgut$ allows us to realize relatively light sleptons in the few hundred GeV range. These sleptons provide additional contributions to muon g-2 that can be arranged to reconcile theory and experiment. The solutions compatible with the muon $g-2$ resolution typically predict light sleptons, charginos and neutralinos. We show how these solutions can be probed during LHC-Run3. A direct impact on the chargino mass is observed such that the chargino can be probed up to about 800 GeV during the Run3 experiments. Despite such a direct impact, it is still possible to realize lighter chargino masses which can escape detection due to the chirality mixture of the lighter slepton states. The colored squarks as well as the gluino turn out to be heavier than about 3-4 TeV if the LSP neutralino satisfies the Planck bound on the dark matter relic abundance. We highlight a variety of benchmark points and, in particular, coannihilation scenarios with dark matter candidates that will be tested in the ongoing and planned direct and indirect detection experiments. 
By relaxing the requirement that the LSP neutralino saturates the relic dark matter abundance, we are able to find solutions with gluino and squark masses in a range that may be accessible at LHC-Run3.
\end{abstract}
\end{titlepage}


\section{Introduction}
\label{sec:intro}

Very recently the FermiLab experiment has provided a new measurement for the muon anomalous magnetic moment (hereafter muon $g-2$ for short), which differs from the Standard Model (SM) prediction by $3.3\sigma$ \cite{Abi:2021gix}. In combination with the earlier measurement at the Brookhaven National Laboratory \cite{Bennett:2006fi}, this now amounts to a $4.2\sigma$ deviation from the SM prediction \cite{Aoyama:2020ynm}, which may be expressed as

\begin{equation}
\Delta a_{\mu} = (25.1 \pm 5.9)\times 10^{-10}
\label{eq:damuexp}
\end{equation}
where $\Delta a_{\mu} \equiv a_{\mu}^{{\rm exp}}-a_{\mu}^{{\rm SM}}$. 

This result leaves a room for new particles which can contribute to muon $g-2$ through direct coupling to the muon \cite{Lu:2021vcp,Chun:2021dwx,Das:2021zea,Cadeddu:2021dqx,Criado:2021qpd,Chakrabarty:2020jro}. In search for this new physics our approach relies on supersymmetry (SUSY) which is motivated, as is well known, by a number of considerations. Among others, this includes a fairly precise unification of the SM gauge couplings assuming TeV scale SUSY, stabilization of the electroweak vacuum \cite{Degrassi:2012ry,Bezrukov:2012sa,Buttazzo:2013uya}, and the presence of a compelling dark matter (DM) candidate assuming that the lightest supersymmetric particle (LSP) is stable. Despite its strong motivations, the collider experiments have brought a strong impact on the spectrum of SUSY particles. The lower mass bound on the colored supersymmetric partners of the SM have steadily increased during the last few years such that the squarks and gluino must be heavier than about 2 TeV or so \cite{Vami:2019slp,Aad:2019ftg}. The mass bounds on gluino decreases to $m_{\tilde{g}}\gtrsim 800$ GeV if it  happens to be the next to LSP (NLSP) in the SUSY spectrum. The lower bounds on the slepton masses including the smuons \cite{CMS:2017fdz,Sirunyan:2017zss,Aaboud:2018jiw} are considerably low in comparison with the colored sparticles. This motivates us to reconsider a class of supersymmetric grand unification theories (GUTs) with non-universal gaugino masses, without disrupting the gauge coupling unification, to address the recently updated muon $g-2$ anomaly. In addition, we assume a flavor symmetry at the grand unification scale ($\mgut$), under which the first two-family matter fields form a doublet, while the third family matter fields are singlet under this flavor symmetry \cite{Babu:2014sga,Babu:2014lwa}. In this framework, the SUSY breaking distinguishes the third family matter fields from those in the first two families so that the third family receives a different mass from the the first two families. We first summarize the salient features of the model, which is followed by a brief discussion of the new contributions to muon $g-2$ provided by SUSY. Even though the squarks and sleptons of the first two families have universal mass at $\mgut$, the renormalization effects allow us to realize suitably heavy colored sparticles, with the corresponding sleptons being considerably lighter, typically in the few hundred GeV range.

The paper is organized as follows: In Section \ref{sec:funmuon} we list the model parameters that are varied in order to accommodate the muon $g-2$ anomaly. The experimental constraints and scanning procedure are outlined in Section \ref{sec:scan}. In Section \ref{sec:spect} we describe the sparticle spectrum that is compatible with the resolution of the muon $g-2$ anomaly, and in Section \ref{sec:Run3} we discuss how this spectrum can be tested at LHC-Run3. The implications for dark matter are discussed in Section \ref{sec:DM}. We also present Tables \ref{tab1} and \ref{tab2} providing a number of benchmark points in this section. Our conclusions are summarized in Section \ref{sec:conc}.

\section{Fundamental Model Parameters and muon $\mathbf{g-2}$}
\label{sec:funmuon}

\begin{figure}[htb!]
\centering
\includegraphics[scale=1.8]{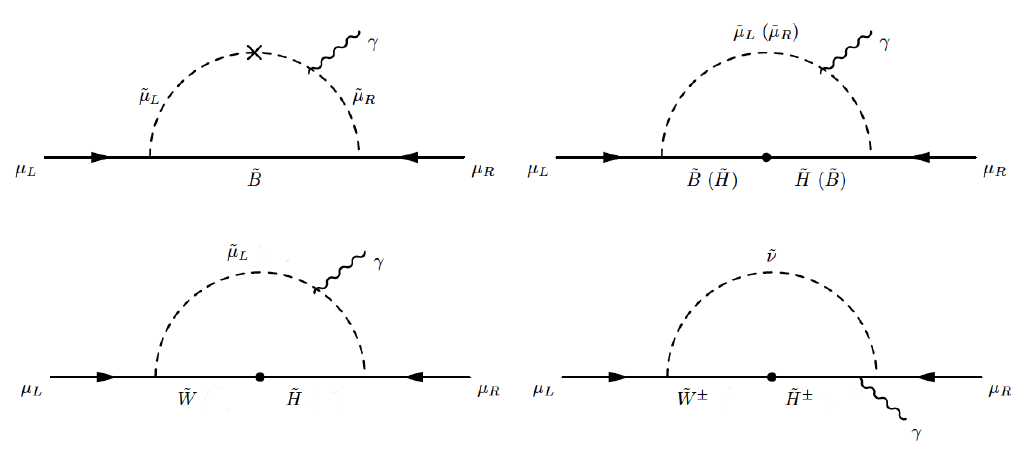}
\caption{The leading contributions to the muon $g-2$ through neutralino and chargino loops. The cross in the top-left diagram denotes the chirality flip between the left- and right-handed smuons, while the dots in the other diagrams represent the mixing between different Neutralino species. In the top-right diagram, there is another loop which is formed by the particles given in the parentheses.}
\label{fig:SNloops}
\end{figure}

The main contributions to muon g-2 from the supersymmetric particles come from the smuon-neutralino and sneutrino-chargino loops. The analytic expressions for these contributions can be formulated by decomposing them into neutralino and chargino species along with the chiral fields of smuons and sneutrinos \cite{Moroi:1995yh,Martin:2001st,Giudice:2012pf}. Figure \ref{fig:SNloops} displays the diagrams for the leading contributions to muon $g-2$ from the neutralino and chargino loops. The cross in the top-left diagram denotes the chirality flip between the left- and right-handed smuons, while the dots in the other diagrams represent the mixing between different neutralino and chargino species. 

As seen from the first diagram (top-left), the chirality flip between the left- and right-handed smuons contributes to muon $g-2$, which is proportional to $\mu\tan\beta - A_{\mu}$, where $A_{\mu}$ is the trilinear SSb coupling between the Higgs doublet and smuons. The contribution from this diagram can be calculated as \cite{Fargnoli:2013zia}: 

\begin{equation}
\Delta a_{\mu}^{\tilde{B}\tilde{\mu}_{L}\tilde{\mu}_{R}} \simeq \dfrac{g_{1}^{2}}{16\pi^{2}}\dfrac{m_{\mu}^{2}M_{\tilde{B}}(\mu\tan\beta - A_{\mu})}{m_{\tilde{\mu}_{L}}^{2}m_{\tilde{\mu}_{R}}^{2}}F_{N}\left( \dfrac{m_{\tilde{\mu}_{L}}^{2}}{M_{\tilde{B}}^{2}},\dfrac{m_{\tilde{\mu}_{R}}^{2}}{M_{\tilde{B}}^{2}}\right)~, \tag{2.1-a}
\label{eq:binocont}
\end{equation}
where

\begin{equation*}
F_{N}(x,y) = xy\left[\dfrac{-3+x +y + xy}{(x-1)^{2}(y-1)^{2}} + \dfrac{2x\log(x)}{(x-y)(x-1)^{3}} -\dfrac{2y\log(y)}{(x-y)(y-1)^{3}}\right]~.
\end{equation*}
Note that there is another diagram in Figure \ref{fig:SNloops}, in which the Bino is replaced with the Higgsino, and the coupling depends on the Yukawa interaction with $y_{\mu}=\mathcal{O}(10^{-3})$. The analytic expression for the Higgsino contribution can be obtained by replacing $M_{\tilde{B}}$ with $\mu$ and $g_{1}$ with $y_{\mu}$ in Eq.(\ref{eq:binocont}). Since $y_{\mu}$ is much smaller than the $U(1)_{Y}$ gauge coupling ($g_{1}$), the Higgsino contribution is usually negligible in comparison with the Bino contribution. However, it is still possible to identify some regions for significant contributions from the Higgsino, where $\mu/M_{\tilde{B}} > (y_{\mu}/g_{1})^{2}$ \cite{Gogoladze:2014cha}.

In addition to the chirality flip between the smuon fields, the mixing among the MSSM neutralinos can also contribute to muon $g-2$ as seen from the other diagrams in Figure \ref{fig:SNloops}. The contributions from the top-right diagram can be formulated as:

\begin{equation}
\setstretch{3}
\begin{array}{ll}
\Delta a_{\mu}^{\tilde{B}\tilde{H}\tilde{\mu}} \simeq  \dfrac{g_{1}^{2}}{16\pi^{2}}m_{\mu}^{2}M_{1}(\mu\tan\beta - A_{\mu}) & \left[
\dfrac{1}{m_{\tilde{\mu}_{L}}^{4}}F_{N}\left(\dfrac{M_{1}^{2}}{m_{\tilde{\mu}_{L}}^{2}}, \dfrac{\mu^{2}}{m_{\tilde{\mu}_{L}}^{2}}   \right) \right.\\
& - \left. \dfrac{2}{m_{\tilde{m}_{R}}^{4}}F_{N}\left(\dfrac{M_{1}^{2}}{m_{\tilde{\mu}_{R}}^{2}}, \dfrac{\mu^{2}}{m_{\tilde{\mu}_{R}}^{2}}   \right) \right]~.
\end{array} \tag{2.1-b}
\end{equation}

There is also contribution from the Wino, which can be obtained by replacing the Bino with Wino as shown in the bottom-left diagram. However, because of the $SU(2)_{L}$ symmetry, only the left-handed smuon can run in the loop for this contribution to be relevant. It can be approximated as:

\begin{equation}
\Delta a_{\mu}^{\tilde{W}\tilde{H}\tilde{\mu}} \simeq  \dfrac{g_{2}^{2}}{16\pi^{2}} \dfrac{m_{\mu}^{2}M_{2}(\mu\tan\beta - A_{\mu})}{m_{\tilde{\mu}_{L}}^{4}}F_{N}\left(\dfrac{M_{2}^{2}}{m_{\tilde{\mu}_{L}}^{2}}, \dfrac{\mu^{2}}{m_{\tilde{\mu}_{L}}^{2}}   \right)~. \tag{2.1-c}
\end{equation}

Finally, the sneutrino also can contribute to muon $g-2$ through loops involving the chargino, as shown in the bottom-right diagrams in Figure \ref{fig:SNloops}. However, due to the absence of the right-handed neutrino, there is no mechanism in MSSM to flip the chirality in the sneutrino line. In order to have a non-zero contribution from such diagrams, there should be considerable mixing between the Wino-like and Higgsino-like charginos. Such contributions can be summarized as:

\begin{equation}
\Delta a_{\mu}^{\tilde{W}\tilde{H}\tilde{\nu}} \simeq \dfrac{g_{2}^{2}}{8\pi^{2}}\dfrac{m_{\mu}^{2}M_{2}(\mu\tan\beta -A_{\mu})}{m_{\tilde{\nu}}^{4}}F_{C}\left(\dfrac{M_{2}^{2}}{m_{\tilde{\nu}}^{2}},\dfrac{\mu^{2}}{m_{\tilde{\nu}}^{2}} \right)~, \tag{2.1-d}
\end{equation}
where

\begin{equation*}
F_{C}(x,y)=xy\left[ \dfrac{5-3(x+y)+xy}{(x-1)^{2}(y-1)^{2}} - \dfrac{2\log(x)}{(x-y)(x-1)^{3}}+\dfrac{2\log(y)}{(x-y)(y-1)^{3}}\right]~.
\end{equation*}
\setcounter{equation}{1}
Putting together the variety of contributions, one can list the fundamental parameters for muon $g-2$ as follows:

\begin{equation*}
M_{\tilde{B}}, M_{\tilde{W}},\mu,\tan\beta,A_{\mu},m_{\tilde{\mu}_{L}},m_{\tilde{\mu}_{R}},m_{\tilde{\nu}_{L}}
\end{equation*}
Among these parameters, the effects of $A_{\mu}$ can usually be neglected since $A_{\mu} \ll \mu\tan\beta$. They can be converted into the GUT scale parameters, namely

\begin{equation}
\begin{array}{ll}
M_{\tilde{B}} \rightarrow M_{1}~, & M_{\tilde{W}} \rightarrow M_{2}~, \\
\mu \rightarrow (m_{H_{d}}, m_{H_{u}})~, & A_{\mu} \rightarrow A_{0}y_{\mu}~, \\
\multicolumn{2}{c}{m_{\tilde{\mu}_{L}},m_{\tilde{\mu}_{R}},m_{\tilde{\nu}} \rightarrow m_{0_{1,2}}~.}
\end{array}
\end{equation}

\section{Experimental Constraints and Scanning Procedure}
\label{sec:scan}

We have performed random scans over the fundamental parameters of the model, which are listed with their ranges as

\begin{equation}
\begin{array}{rll}
0 \leq & m_{0_{1,2}}, m_{0_{3}} & \geq 5~{\rm TeV}, \\
0 \leq & M_{1},M_{2} & \geq 2~{\rm TeV}, \\
-5 \leq & M_{3} & \geq 5~{\rm TeV}, \\
-3 \leq & A_{0}/m_{0_{3}} & \geq 3, \\
1.2 \leq & \tan\beta & \leq 60 \\
0 \leq & m_{H_{d}}, m_{H_{u}} & \leq 5~{\rm TeV}~,
\end{array}
\end{equation}
where the SSB masses for the matter fields are set at $\mgut$ with $m_{0_{1,2}}$ for the first two-family fields, and with $m_{0_{3}}$ for the third family matter fields. This set up is configured by assuming a $SO(3)$ flavor symmetry at $\mgut$ under which the first two families form a doublet, and the third family matter fields transform as a singlet. The SSB gaugino masses are parametrized as $M_{1}, M_{2}$ for the $U(1)_{Y}$ and $SU(2)_{L}$ gauginos, and $M_{3}$ determines the gluino mass. The trilinear scalar coupling is determined from its ratio to $m_{0_{3}}$, and it is restricted to be less than 3 to avoid the color/charge breaking minima \cite{Ellwanger:1999bv}. The MSSM parameter $\tan\beta \equiv \langle H_{u}\rangle / \langle H_{d} \rangle$ is the ratio of VEVs of the MSSM Higgs doublets, as previously defined, and $m_{H_{d}}, m_{H_{u}}$ are the SSB masses of the MSSM Higs fields.

We employ SPheno-4.0.4 \cite{Porod:2003um,Porod:2011nf} in our scans which is generated by SARAH-4.14.4 \cite{Staub:2008uz,Staub:2015iza}. This package first runs the renormalization group equations (RGEs) of the effective Yukawa couplings and the gauge couplings of the SM from $M_{Z}$, and determines the scale ($\mgut$) at which the gauge couplings unify. Note that, we do not impose strict unification at $\mgut$, since $g_{3}$ can receive some threshold corrections from some unknown symmetry breaking mechanism around $\mgut$. Thus the unification scale is determined by the condition $g_{1}=g_{2}\simeq g_{3}$, in which deviation from unification of $g_{3}$ is restricted up to $3\%$. After the unification scale is determined, the RGEs are run back to $M_{Z}$ together with the SSB terms to calculate the low scale mass spectrum, decay channels and observables.

After we generate the data, we employ some theoretical and experimental constraints in our analyses. The radiative electroweak symmetry breaking (REWSB) constraint is one of the leading ones \cite{Ibanez:1982fr,Inoue:1982pi,Ibanez:1982ee,Ellis:1982wr,AlvarezGaume:1983gj}. The $\mu-$term is calculated by inputting the low scale values of $m_{H_{d}}$ and $m_{H_{d}}$ into the tadpole equations, but its sign remains indefinite. We assume the sign to be positive for all solutions. Another strong constraint arises from considerations of dark matter, and we only retain those solutions which yield one of the MSSM neutralinos as the LSP. The current Planck bound \cite{Akrami:2018vks} on the relic abundance of LSP neutralino still brings another strong constraint. On the other hand, even though a solution does not yield a correct relic density, it can still survive in conjunction with other form(s) of dark matter \cite{Li:2014xqa,Baer:2012by}. In such cases, LSP neutralino still takes part in dark matter, but it is not accounted for all the observations. We should note that this case still excludes solutions if they yield relic abundance of LSP neutralino larger than Planck bound. In this context, we also consider the solutions in the low relic density regions.

In addition to these constraints we successively apply the mass bounds on supersymmetric particles \cite{Agashe:2014kda} and the constraints from rare $B-$meson decays such as $B_{s}\rightarrow \mu^{+}\mu^{-}$ \cite{Aaij:2012nna} and $B_{s}\rightarrow X_{s}\gamma$ \cite{Amhis:2012bh}. The branching ratios of these decay modes are calculated by FlavorKit \cite{Vicente:2014xda} interfaced into the SPheno package. Even though we do not apply constraints from other rare $B-$meson decays, we require solutions that perform no worse than the SM predictions.

In applying the mass bounds, we list only the Higgs and gluino mass bounds, since they have been updated by the LHC experiments. Below is the summary of the experimental constraints employed in our analyses:

\begin{equation}
\setstretch{1.8}
\begin{array}{c}
m_h  = 123-127~{\rm GeV}\\
m_{\tilde{g}} \geq 2.1~{\rm TeV}\\
0.8\times 10^{-9} \leq{\rm BR}(B_s \rightarrow \mu^+ \mu^-) \leq 6.2 \times10^{-9} \;(2\sigma) \\
2.99 \times 10^{-4} \leq  {\rm BR}(B \rightarrow X_{s} \gamma)  \leq 3.87 \times 10^{-4} \; (2\sigma) \\
0.114 \leq \Omega_{{\rm CDM}}h^{2} \leq 0.126~.
\label{constraints}
\end{array}
\end{equation}

\section{Supersymmetric Mass Spectrum compatible with Muon $\mathbf{g-2}$}
\label{sec:spect}

\begin{figure}[ht!]
\centering
\subfigure{\includegraphics[scale=0.4]{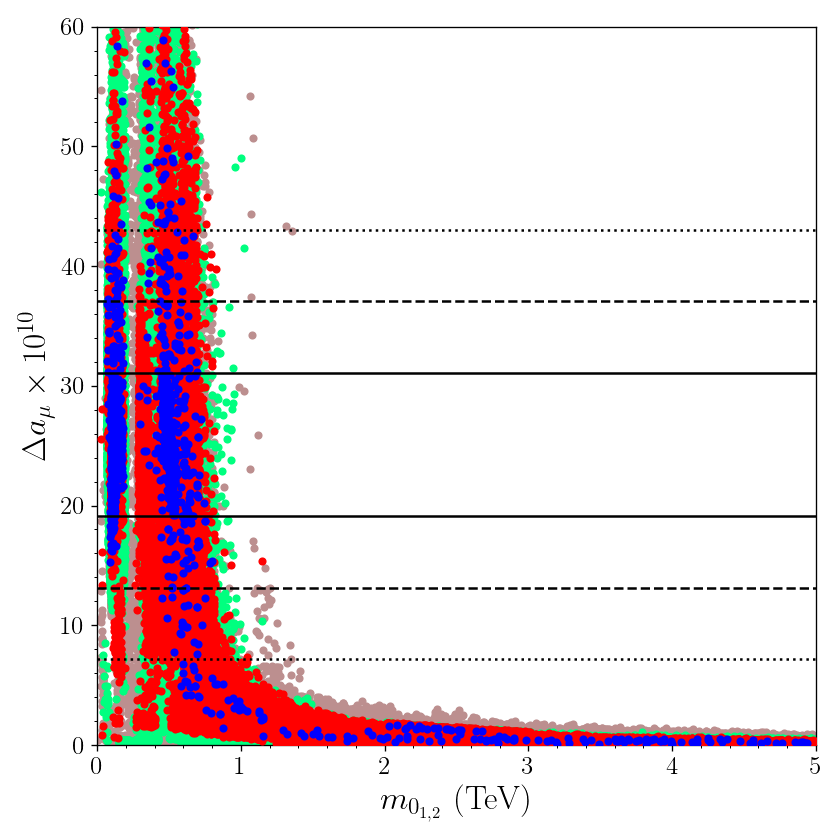}}%
\subfigure{\includegraphics[scale=0.4]{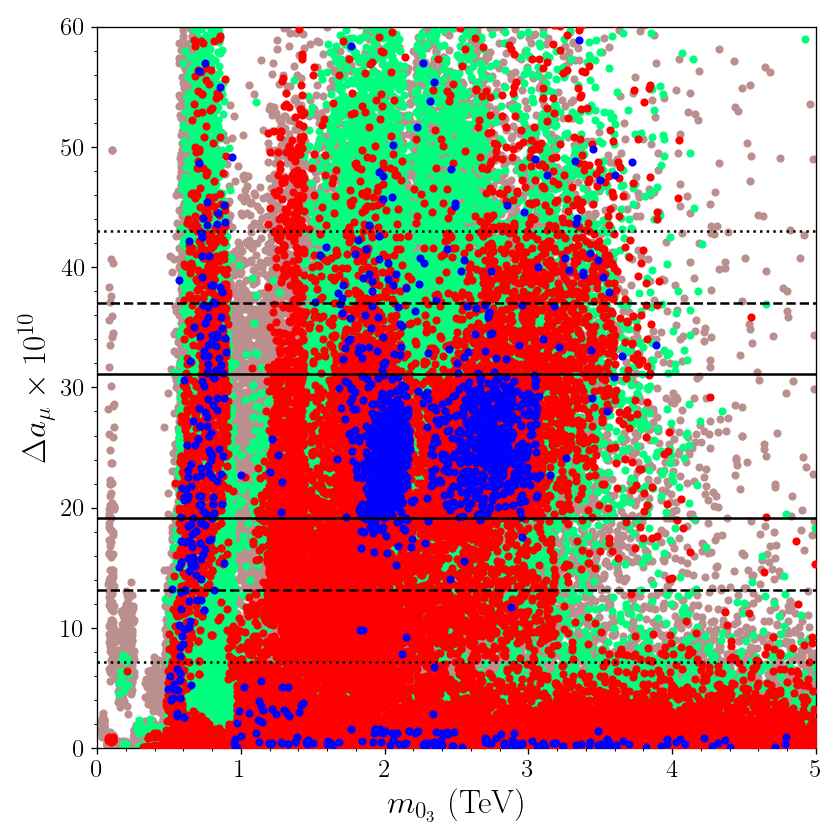}}
\subfigure{\includegraphics[scale=0.4]{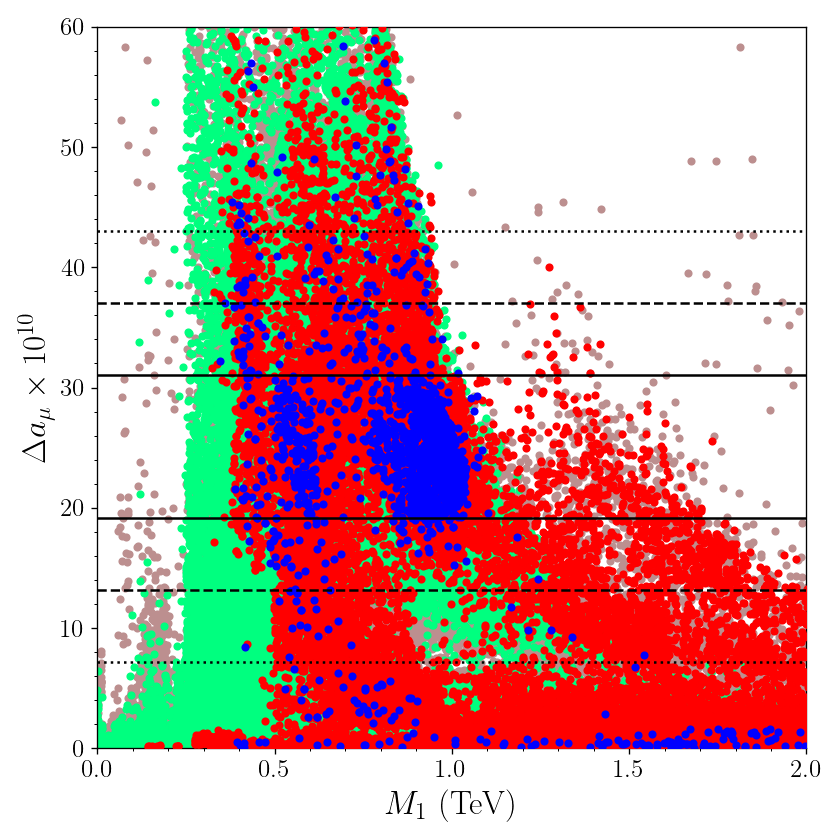}}%
\subfigure{\includegraphics[scale=0.4]{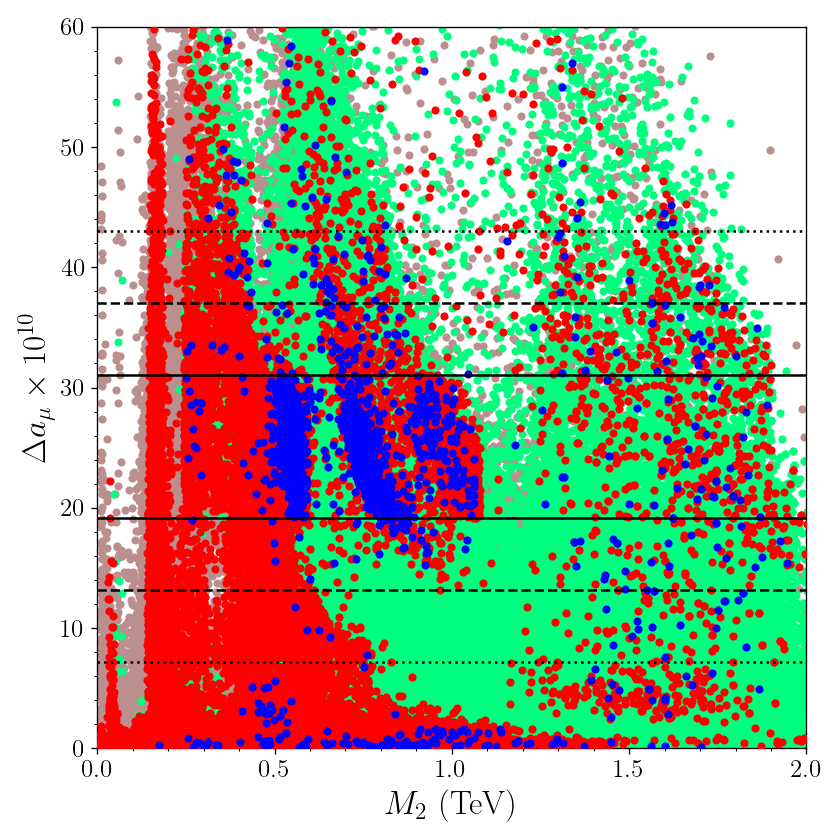}}
\caption{Plots in the $\Delta a_{\mu}-m_{0_{1,2}}$, $\Delta a_{\mu}-m_{0_{3}}$, $\Delta a_{\mu}-M_{1}$ and $\Delta a_{\mu}-M_{2}$ planes. All points are compatible with the REWSB and LSP neutralino conditions. Green points satisfy the mass bounds and the constraints from rare $B-$meson decays. Blue and red points form subsets of green. The solutions consistent with the Planck bound within $5\sigma$ uncertainty are shown in blue, while those yielding low relic density are represented in red. The solid, dashed and horizontal lines show the regions which resolve the muon $g-2$ anomaly within the $1\sigma$, $2\sigma$ and $3\sigma$ intervals.}
\label{fig2}
\end{figure}

\begin{figure}[ht!]
\centering
\subfigure{\includegraphics[scale=0.4]{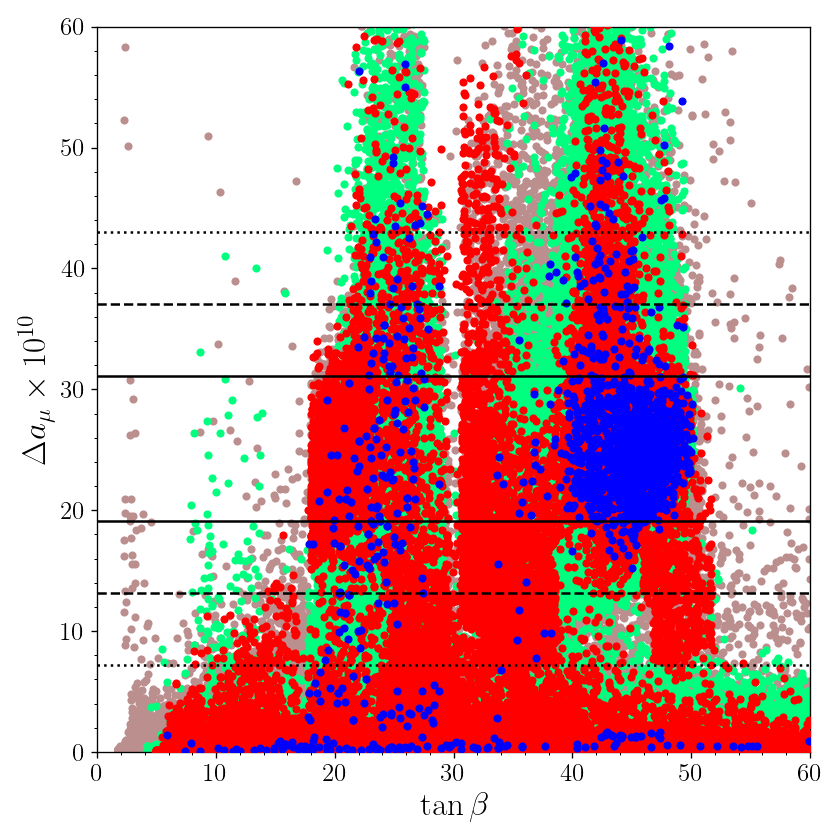}}%
\subfigure{\includegraphics[scale=0.4]{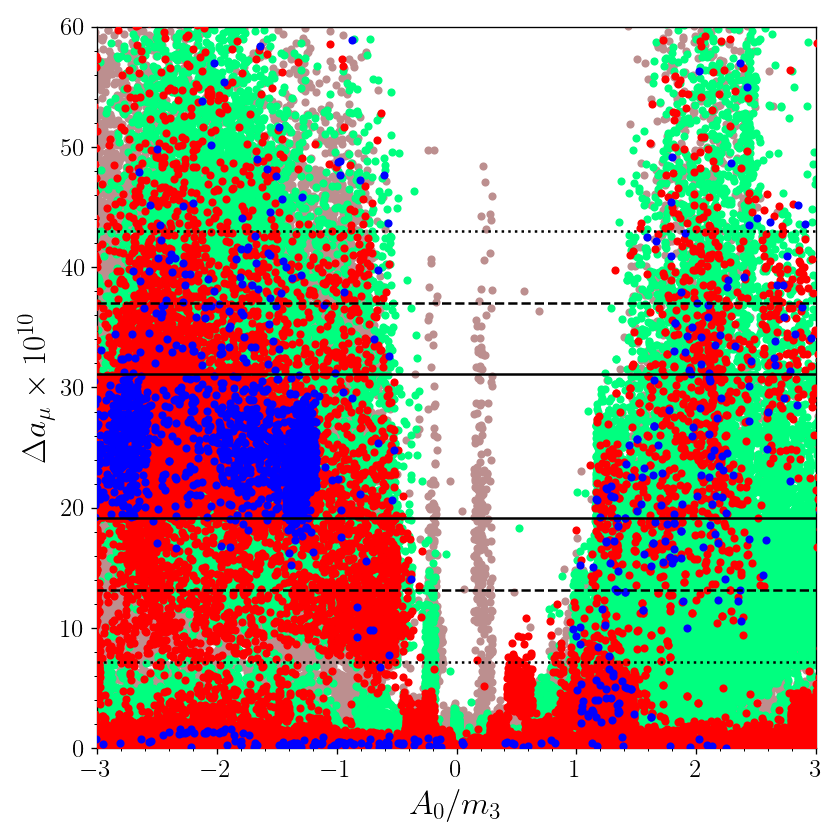}}
\caption{Plots in the $\Delta a_{\mu}-\tan\beta$ and $\Delta a_{\mu}-A_{0}/m_{3}$ planes. The color coding is the same as in Figure \ref{fig2}.}
\label{fig3}
\end{figure}

\begin{figure}[ht!]
\centering
\subfigure{\includegraphics[scale=0.4]{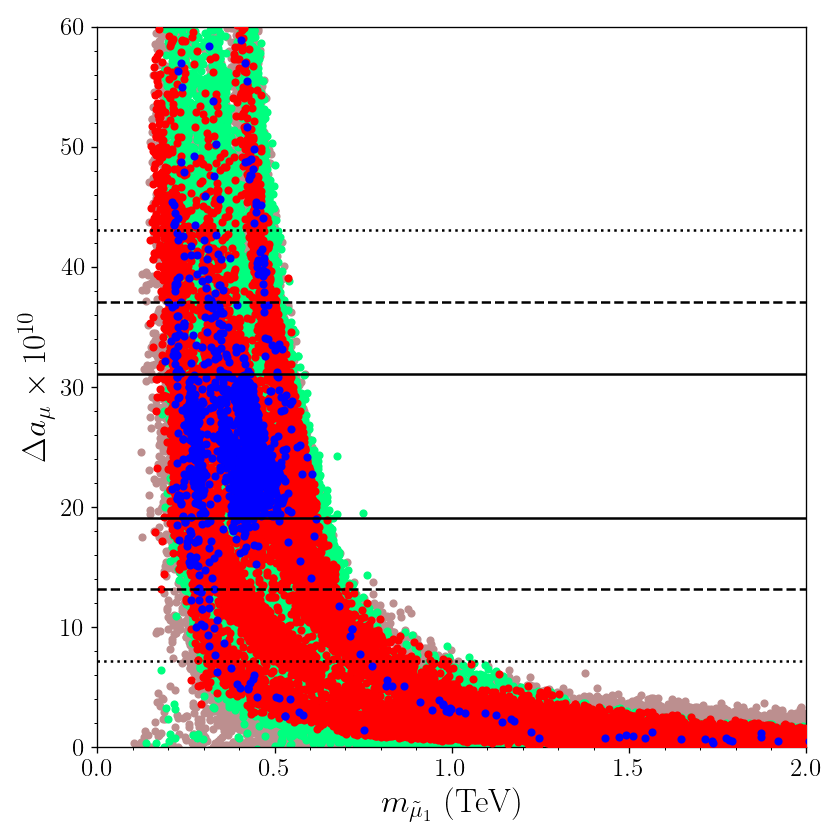}}%
\subfigure{\includegraphics[scale=0.4]{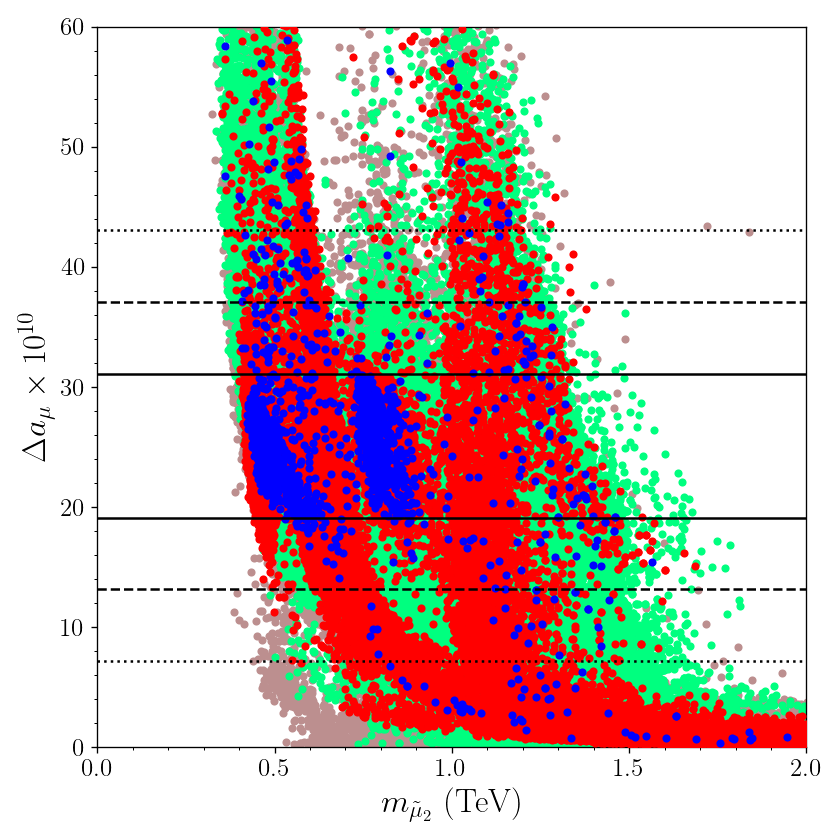}}
\subfigure{\includegraphics[scale=0.4]{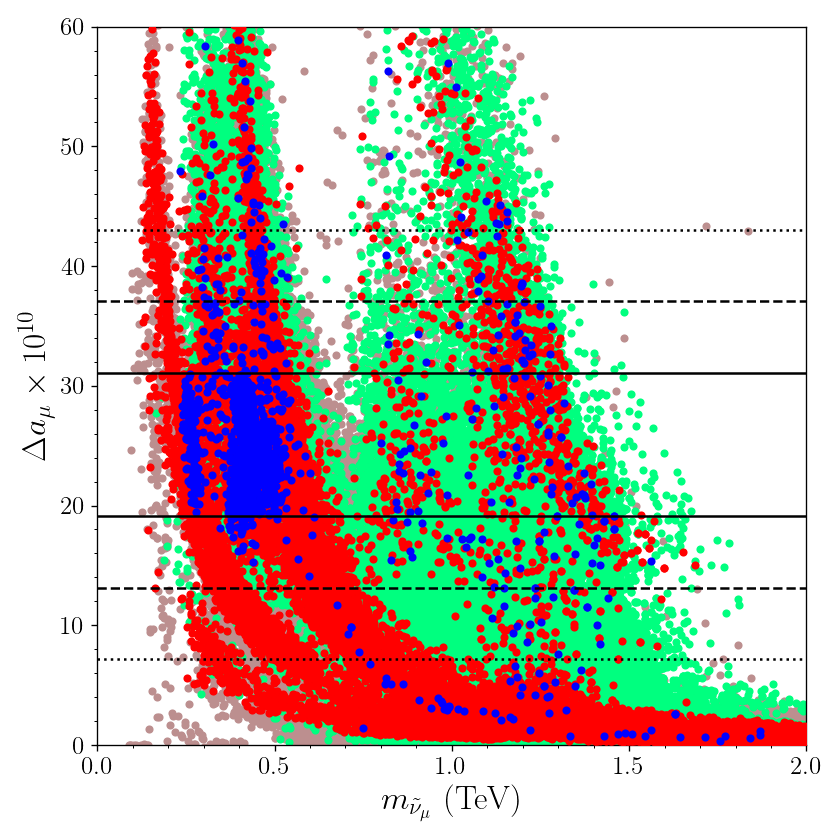}}%
\subfigure{\includegraphics[scale=0.4]{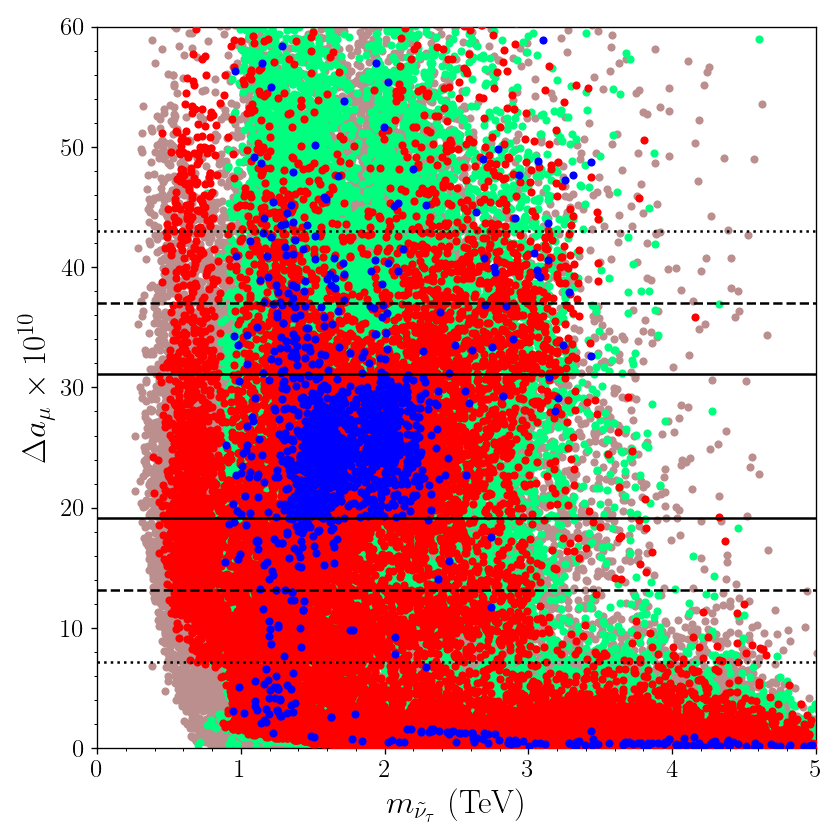}}
\caption{Smuon and sneutrino masses with plots in the $\Delta a_{\mu}-m_{\tilde{\mu}_{1}}$, $\Delta a_{\mu}-m_{\tilde{\mu}_{2}}$, $\Delta a_{\mu}-m_{\tilde{\nu}_{\mu}}$ and $\Delta a_{\mu}-m_{\tilde{\nu}_{\tau}}$ planes. The color coding is the same as in Figure \ref{fig2}.}
\label{fig4}
\end{figure}

\begin{figure}[ht!]
\centering
\subfigure{\includegraphics[scale=0.4]{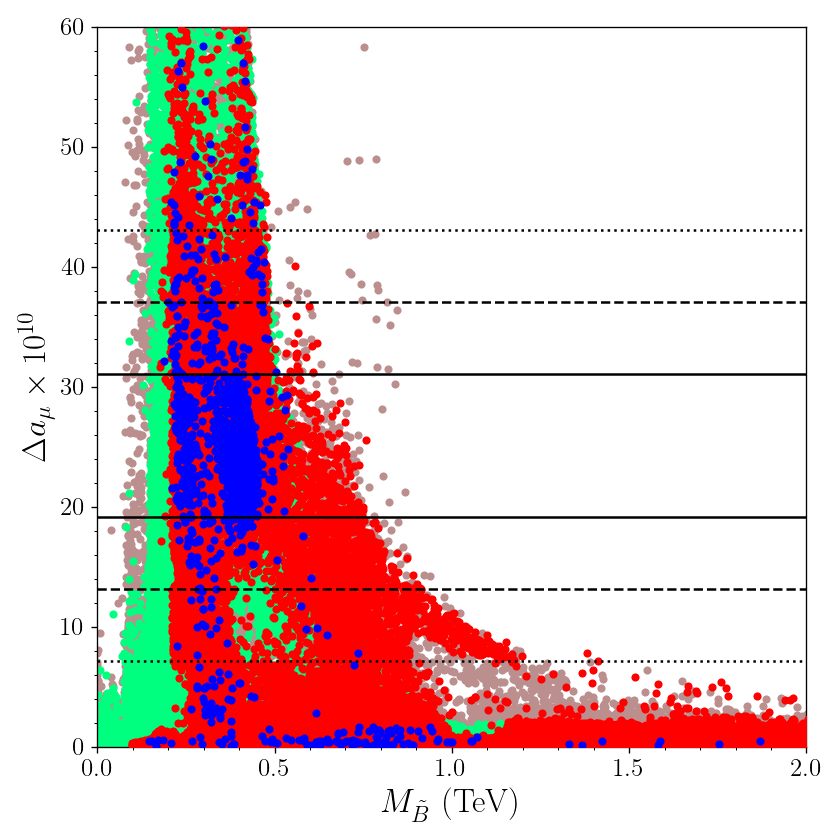}}%
\subfigure{\includegraphics[scale=0.4]{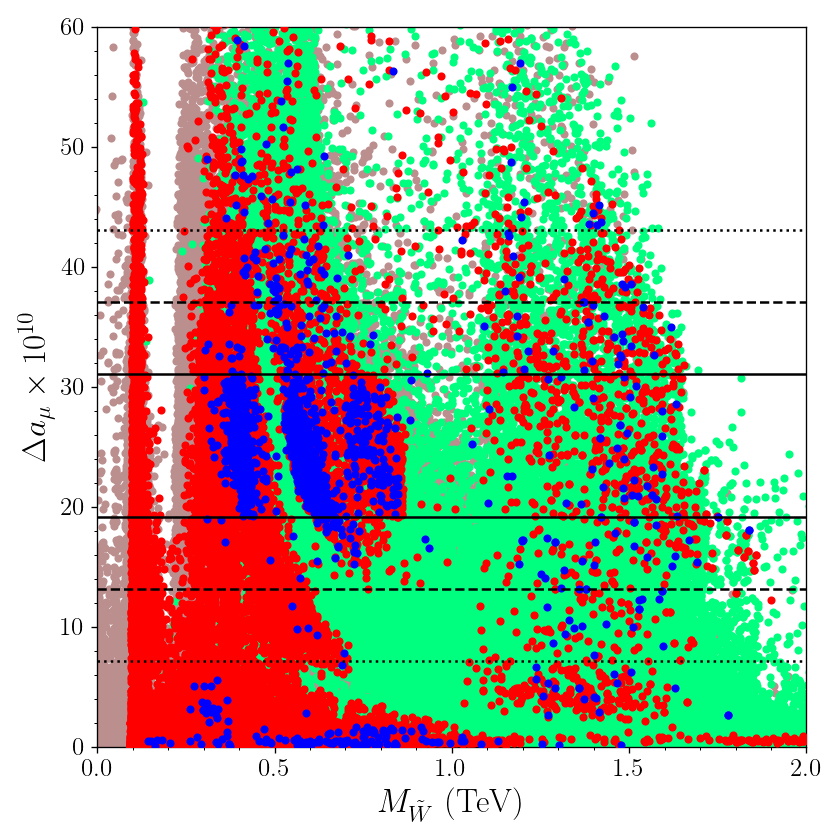}}
\subfigure{\includegraphics[scale=0.4]{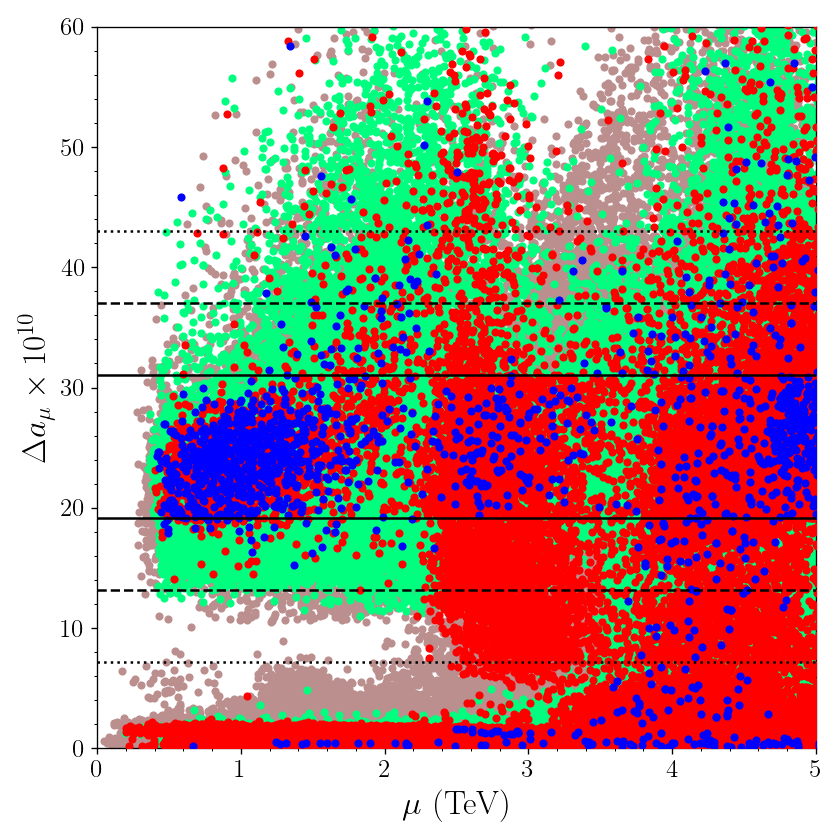}}%
\subfigure{\includegraphics[scale=0.4]{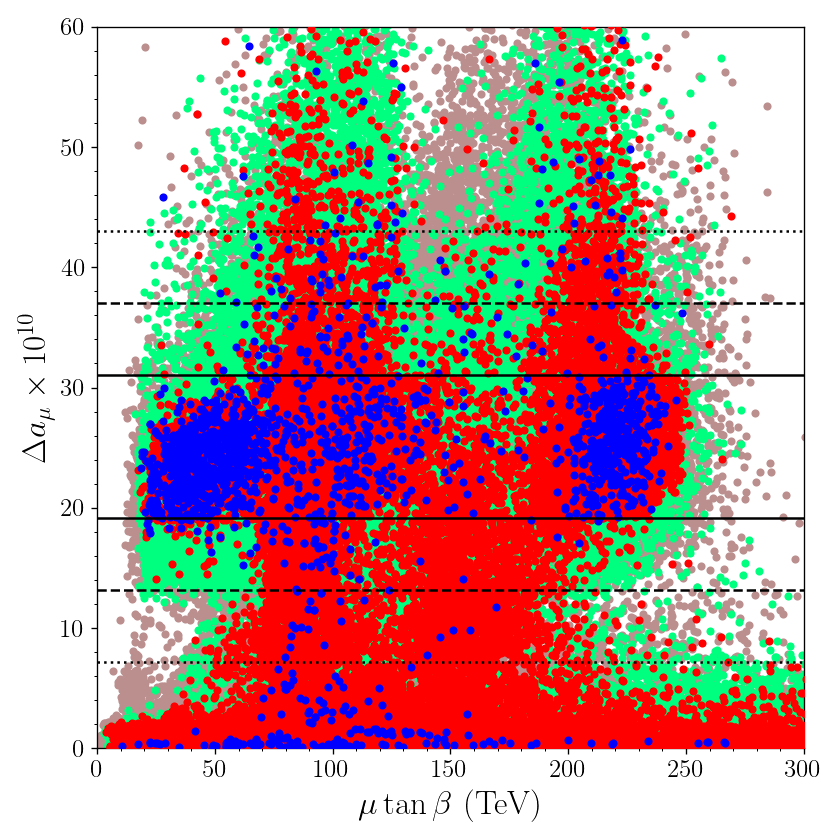}}
\caption{Smuon and sneutrino masses with plots in the $\Delta a_{\mu}-M_{\tilde{B}}$, $\Delta a_{\mu}-M_{\tilde{W}}$, $\Delta a_{\mu}-\mu$ and $\Delta a_{\mu}-\mu\tan\beta$ planes. The color coding is the same as in Figure \ref{fig2}.}
\label{fig5}
\end{figure}

\begin{figure}[ht!]
\centering
\subfigure{\includegraphics[scale=0.4]{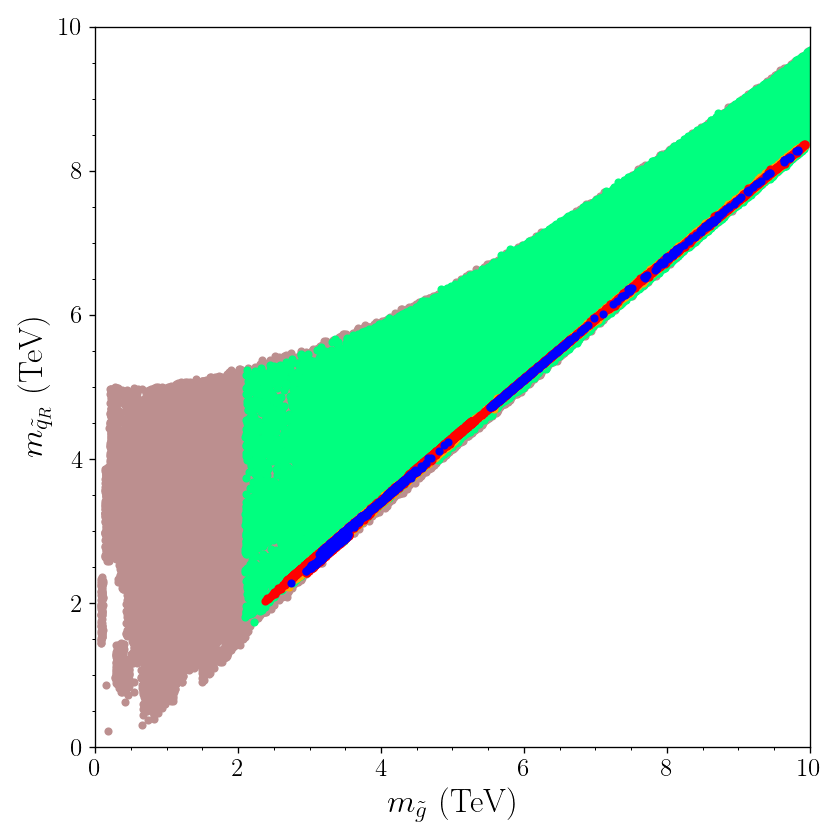}}%
\subfigure{\includegraphics[scale=0.4]{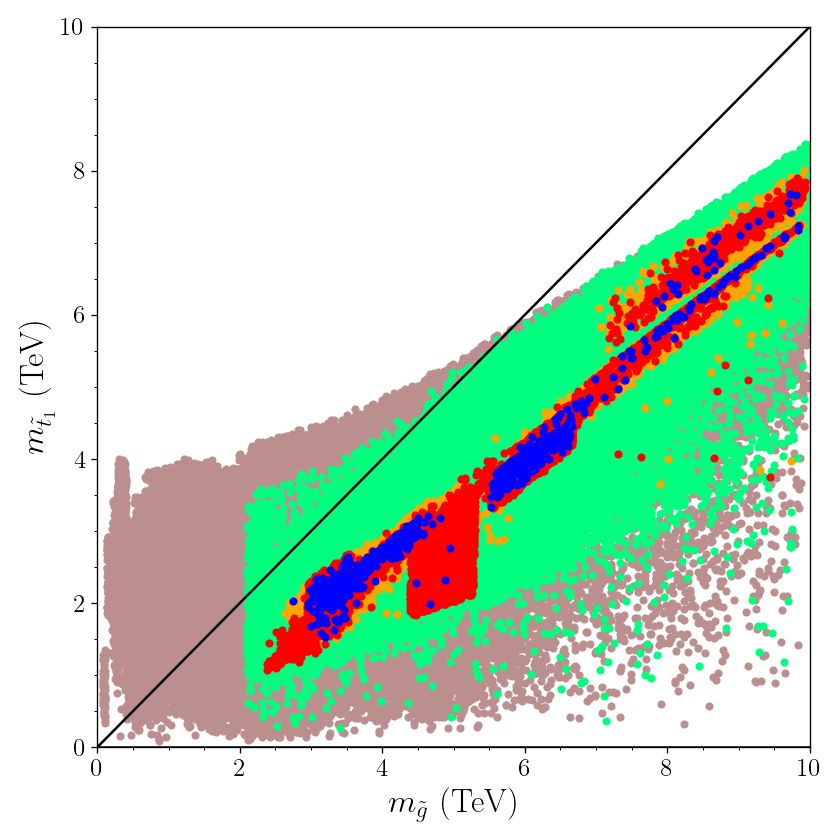}}
\caption{Plots in the  $m_{\tilde{q}_{R}}-m_{\tilde{g}}$ and $m_{\tilde{t}_{1}}-m_{\tilde{g}}$ planes. All points are compatible with the REWSB and LSP neutralino conditions. Green points satisfy the mass bounds and constraints from rare $B-$meson decays. Orange points form a subset and they provide a resolution to the muon $g-2$ discrepancy within $1\sigma$. Red and blue points form subsets of orange. The solutions leading to a relic density for the LSP neutralino consistent with the Planck bound within $5\sigma$ are shown in blue, and those with low relic density are represented in red. The diagonal line in the $m_{\tilde{t}_{1}}-m_{\tilde{g}}$ plane shows the regions where $m_{\tilde{t}_{1}}=m_{\tilde{g}}$.}
\label{fig6}
\end{figure}

In this section, we discuss the impact from the resolution to the muon $g-2$ discrepancy within $1\sigma$ on the mass spectrum of the supersymmetric particles. Figure \ref{fig2} displays our results for the fundamental parameters, which are imposed at $\mgut$ in correlation with the muon $g-2$ results with plots in the $\Delta a_{\mu}-m_{0_{1,2}}$, $\Delta a_{\mu}-m_{0_{3}}$, $\Delta a_{\mu}-M_{1}$ and $\Delta a_{\mu}-M_{2}$ planes. All points are compatible with the REWSB and LSP neutralino conditions. Green points satisfy the mass bounds and the constraints from rare $B-$meson decays. Blue and red points form subsets of green. The solutions consistent with the Planck bound within $5\sigma$ uncertainty are shown in blue, while those yielding low relic density are represented in red. The solid, dashed and horizontal lines show the regions which resolve the muon $g-2$ anomaly within the $1\sigma$, $2\sigma$ and $3\sigma$ intervals. The top panels of Figure \ref{fig2} show the SSB mass terms for the matter fields and as expected, the current experimental measurements of $\Delta a_{\mu}$ show a strong impact on the masses $m_{0_{1,2}}$ of the first two families. A resolution of the muon $g-2$ anomaly up to $2\sigma$ interval restricts $m_{0_{1,2}}$ to be lighter than about 1 TeV. However, extending the resolution interval to $3\sigma$ leads to a slight relaxation on $m_{0_{1,2}}$. The results in the $\Delta a_{\mu}-m_{0_{1,2}}$ plane show that $m_{0_{1,2}}$ should be lighter than about 1.2 TeV to accommodate the muon g-2 solutions in conjunction with the other constraints in MSSM. With the first two-family sfermions relatively light, the third family mass term can lie in a wide range between 0.2 and 5 TeV, as seen from the $\Delta a_{\mu}-m_{0_{3}}$. 

The bottom planes of Figure \ref{fig2} display the SSB mass terms for the $U(1)_{Y}$ (left) and $SU(2)_{L}$ (right) gauginos. In contrast to $m_{0_{1,2}}$, it is possible to realize the muon $g-2$ solutions , even within $1\sigma$, when the electroweakinos are as heavy as about 2 TeV. On the other hand, if one requires the solutions to yield the correct relic density for LSP neutralino, then the Planck measurements bounds $M_{1}$ at about 1 TeV from above to accommodate the muon $g-2$ solutions within $1\sigma$. However, this bound can be shifted up to about 1.6 TeV if one extends the interval for the muon $g-2$ solutions to $3\sigma$ as seen from the $\Delta a_{\mu}-M_{1}$ plane. Similarly, even though the muon $g-2$ solutions favor the regions with light $M_{2}$, it is still possible to realize the muon $g-2$ resolution within $1\sigma$ when $M_{2} \lesssim 2$ TeV. The $\Delta a_{\mu}-M_{2}$ plane shows that $M_{2}$ lies between about 300 GeV and 2 TeV.. 

Figure \ref{fig3} shows our results for the other relevant parameters for supersymmetric contributions to the muon $g-2$ with plots in the $\Delta a_{\mu}-\tan\beta$ and $\Delta a_{\mu}-A_{0}/m_{3}$ planes. The color coding is the same as in Figure \ref{fig2}. The $\Delta a_{\mu}-\tan\beta$ plane shows that the resolutions to the muon $g-2$ discrepancy better than $3\sigma$ requires $\tan\beta \gtrsim 10$ (red), and the Planck bound within $5\sigma$ extends this bound to $\tan\beta \gtrsim 20$ (blue). As discussed before, for the compatible ranges of $\tan\beta$ shown in Figure \ref{fig3}, the effects from the $A_{\mu}$ term becomes negligible, and as seen from the $\Delta a_{\mu}-A_{0}/m_{3}$ plane, the solutions can be realized for any $A_{0}$ value, and most of the solutions accumulate in the region $A_{0}  < 0$. Even though its contribution is relatively small, $A_{0}$ contributes positively to muon $g-2$ in this region and helps improve the results.

The parameters discussed above are the main ones shaping the mass spectrum of supersymmetric particles. We first discuss the masses of the supersymmetric partners of the muon and neutrinos with plots in the $\Delta a_{\mu}-m_{\tilde{\mu}_{1}}$, $\Delta a_{\mu}-m_{\tilde{\mu}_{2}}$, $\Delta a_{\mu}-m_{\tilde{\nu}_{\mu}}$ and $\Delta a_{\mu}-m_{\tilde{\nu}_{\tau}}$ planes in Figure \ref{fig4}. The color coding is the same as in Figure \ref{fig2}. The sharpest impact from the muon $g-2$ solutions is observed in the mass of the lighter smuon state. As seen from the $\Delta a_{\mu}-m_{\tilde{\mu}_{1}}$ plane, the compatible mass scales for $\tilde{\mu}_{1}$ exhibit a behavior quite similar to that of $m_{0_{1,2}}$. The resolution to muon $g-2$ up to $2\sigma$ interval requires $m_{\tilde{\mu}_{1}} \lesssim 800$ GeV, and it can be raised up to about 1.1 TeV if one requires muon $g-2$ solutions within a $3\sigma$ interval. Even though the heavier smuon state can be as heavy as about 1.7 TeV, a lighter $\tilde{\mu}_{2}$ is also favorable for the muon $g-2$ solutions, as shown in the $\Delta a_{\mu}-m_{\tilde{\mu}_{2}}$ plane. These bounds are consistent with both the desired or lower relic density of LSP neutralino as shown in blue and red, respectively. 

The muon $g-2$ condition also has some impact on the masses of sneutrinos, and the muon-sneutrino can be as heavy as about 1.5 TeV in the regions corresponding to  muon $g-2$ solutions. Since the sneutrino loops contributing to muon $g-2$ involves couplings to Higgsinos ($y_{\mu} \sim 10^{-3}$), the sneutrinos can provide only minor contributions. We also display the mass of the tau-sneutrino $\tilde{\nu}_{\tau}$ in the $\Delta a_{\mu}-m_{\tilde{\nu}_{\tau}}$ plane. Since the MSSM does not provide any mixing mechanism in the sneutrinos, $\tilde{\nu}_{\tau}$ does not contribute to muon $g-2$, and so there is no observed impact on its mass which can lie anywhere from about 500 GeV to 5 TeV.On the other hand, the current Planck measurements can provide a bound as $m_{\tilde{\nu}_{\tau}} \gtrsim 1$ TeV (blue).

A similar impact from muon $g-2$ can also be observed in the masses of the MSSM neutralinos. We display our results in Figure \ref{fig5} with plots in the $\Delta a_{\mu}-M_{\tilde{B}}$, $\Delta a_{\mu}-M_{\tilde{W}}$, $\Delta a_{\mu}-\mu$ and $\Delta a_{\mu}-\mu\tan\beta$ planes. The color coding is the same as in Figure \ref{fig2}. The strongest impact is observed on the Bino mass as seen from the $\Delta a_{\mu}-M_{\tilde{B}}$ plane. The Bino cannot be heavier than about 1.3 TeV to accommodate the muon $g-2$ solutions consistent with the mass bounds and $B-$meson decays, even though the solutions yield a low relic abundance of LSP neutralino in this region. If one requires the solutions such that the dark matter should be saturated only by the LSP neutralino, the solutions (blue) reveal an upper bound $M_{\tilde{B}} \lesssim 600$ GeV. Such an upper bound on the Bino mass leads it to play a significant role in the composition of the LSP neutralino. Similarly, the light Wino solutions are favorable for resolving the muon $g-2$ problem, but it is not necessary since also heavy Wino solutions are realized, as shown in the $\Delta a_{\mu}-M_{\tilde{W}}$ plane. The $\Delta a_{\mu}-\mu$ plane shows that muon $g-2$ favors a relatively large $\mu-$term, which results in heavy Higgsinos in the mass spectrum. However, it is still possible to realize solutions with $\mu \gtrsim 500$ GeV. Comparing these results with those shown in the $\Delta a_{\mu}-\mu\tan\beta$ plane, we may conclude that the low $\mu-$term solutions require relatively large $\tan\beta$ ($\gtrsim 40$) to accommodate muon $g-2$.

We can conclude our discussion on the mass spectrum by including the strongly interacting supersymmetric particles as displayed in Figure \ref{fig6} with plots in the  $m_{\tilde{q}_{R}}-m_{\tilde{g}}$ and $m_{\tilde{t}_{1}}-m_{\tilde{g}}$ planes. All points are compatible with the REWSB and LSP neutralino conditions. Green points satisfy the mass bounds and constraints from rare $B-$meson decays. Orange points form a subset and they provide a resolution to the muon $g-2$ discrepancy within $1\sigma$. Red and blue points form subsets of orange. The solutions leading to a relic density for the LSP neutralino consistent with the Planck bound within $5\sigma$ are shown in blue, while those with low relic density are represented in red. Even though they do not contribute to muon $g-2$, $m_{0_{1,2}}$ also determine the masses of the squarks from the first two families, and the impact of muon $g-2$ on the smuon masses may lead to lighter squarks in the spectrum \cite{Babu:2014lwa}. This correlation between masses may be circumvented through a choice of the sign of $M_{3}$ in the RG evolution of the squarks. As a result, as shown in the $m_{\tilde{q}_{R}}-m_{\tilde{g}}$ plane, squark masses can be realized in 2-8 TeV range. The solutions within this whole mass scale are compatible with the muon $g-2$ solution within the $1\sigma$ interval. Even though the SUSY spectra can include squarks lighter than 2 TeV, these solutions are excluded by the mass bound on gluinos. We plot the right-handed squark masses of the first two families denoted with $\tilde{q}_{R}$, since the lighter squark mass eigenstate is mostly formed by the right-handed squarks.

We also discuss the stop masses in comparison with the gluino mass in the $m_{\tilde{t}_{1}}-m_{\tilde{g}}$ plane of Figure \ref{fig6}. The stop masses can be as light as about 1 TeV, which has already been probed at the current LHC experiments depending on the decay channels of stop and chargino \cite{Mitrevski:2018iyy}, and these bounds are applicable over the fundamental parameter space of MSSM. LHC-Run3 is expected to probe the stop mass up to about 1.4 TeV  \cite{Altin:2020qmu}, and the High Luminosity LHC (HL-LHC) can extend this to about 2 TeV.  The diagonal line in this plane shows the regions where $m_{\tilde{t}_{1}}=m_{\tilde{g}}$ and as shown, all muon $g-2$ solutions compatible with the relic density conditions (red and blue) remain under the diagonal line, which indicates $m_{\tilde{g}} \gtrsim m_{\tilde{t}_{1}}$. This region can also be probed through the gluino decays into stops together with the top-quarks. Run3 can slightly improve the gluino probe to extend the mass scales to about 2.3 TeV; on the other hand, this channel can indirectly probe stops up to about 1.9 TeV \cite{Altin:2019veq}, and HL-LHC can extend the stop probe scale to about 3 TeV. 

\section{Probing muon $\mathbf{g-2}$ compatible spectra at LHC-Run3}
\label{sec:Run3}

\begin{figure}[ht!]
\centering
\subfigure{\includegraphics[scale=0.4]{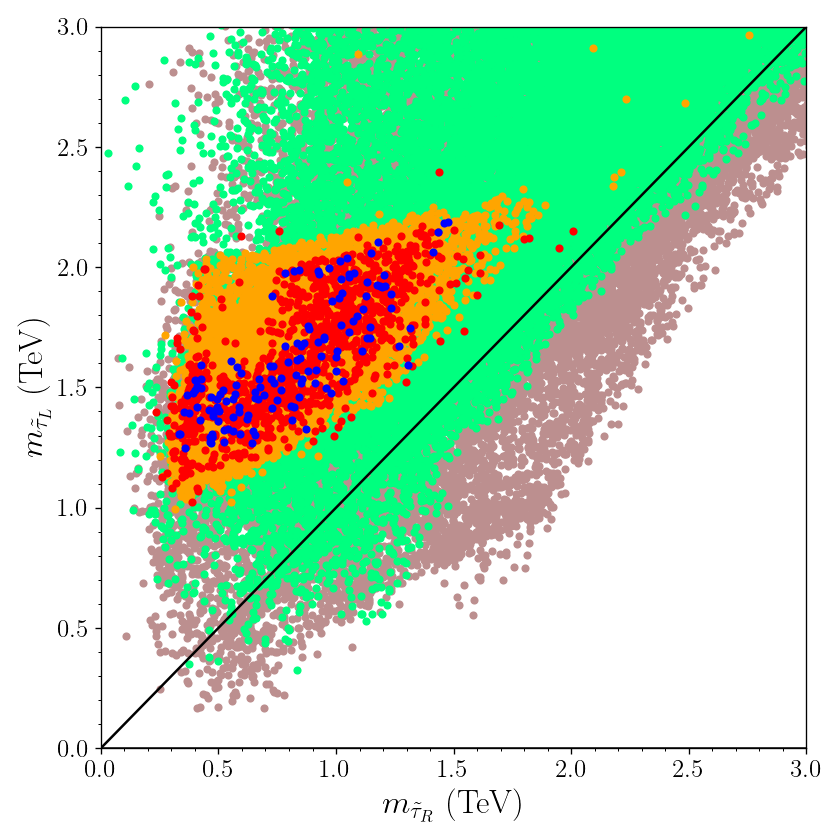}}%
\subfigure{\includegraphics[scale=0.4]{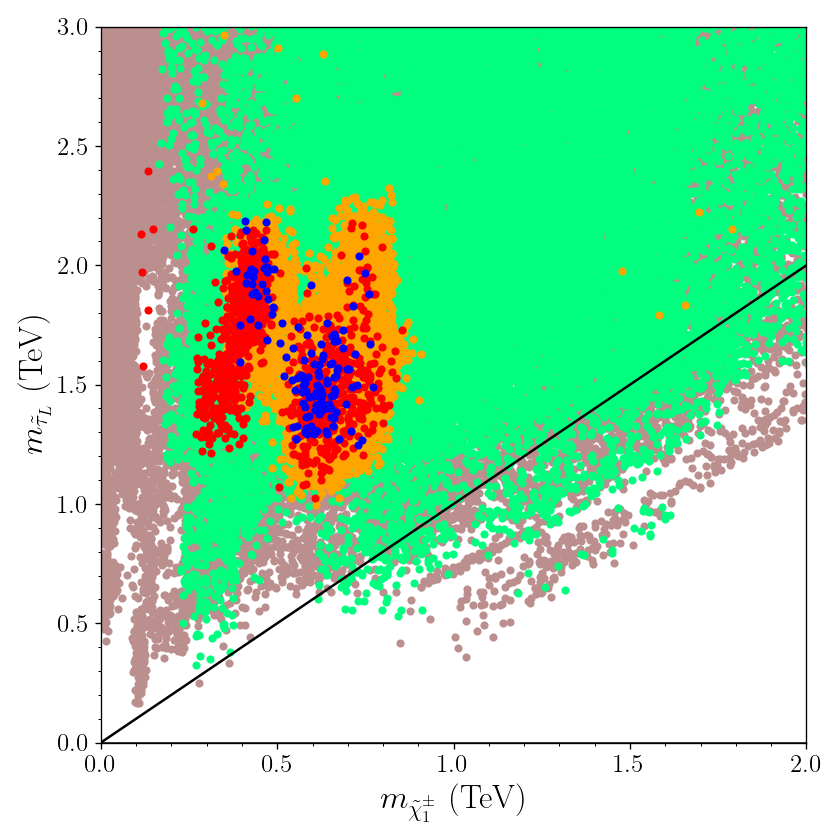}}
\subfigure{\includegraphics[scale=0.4]{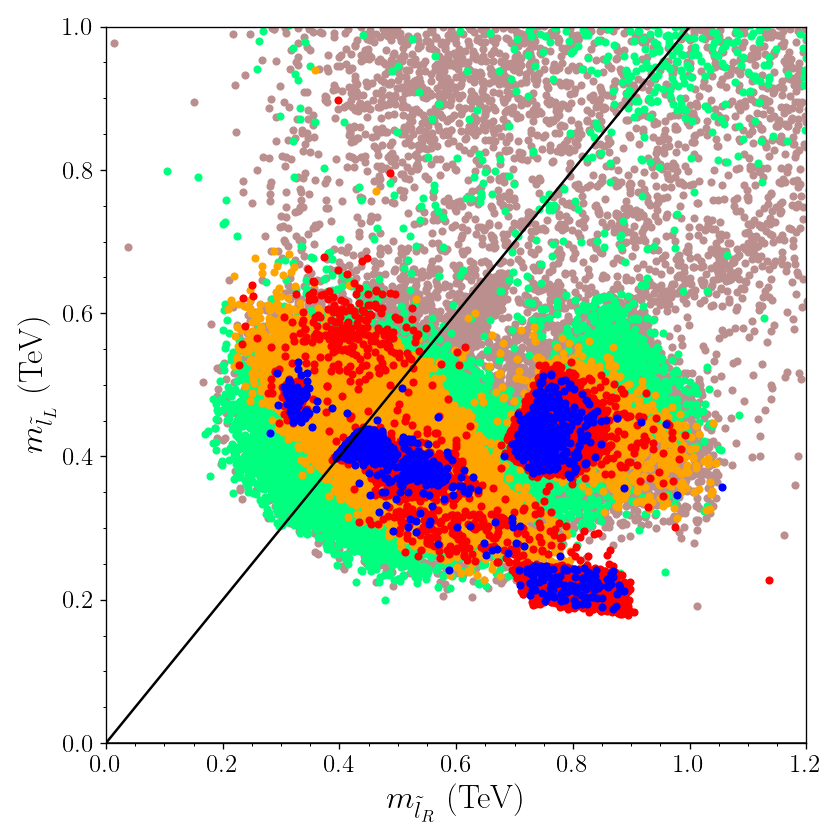}}%
\subfigure{\includegraphics[scale=0.4]{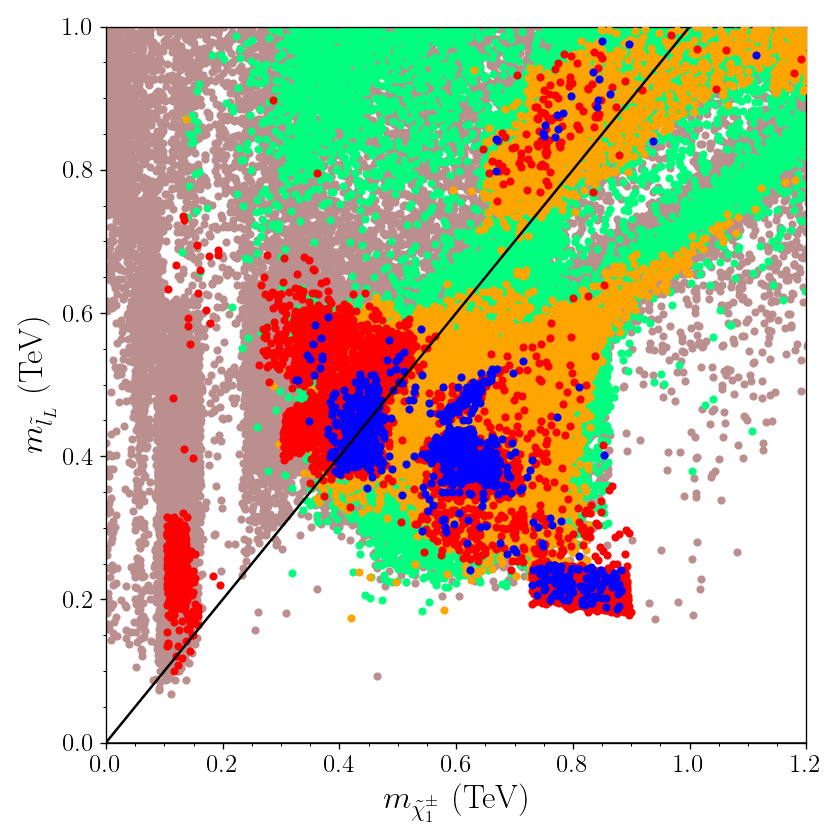}}
\caption{Results for the slepton and chargino masses with plots in the $m_{\tilde{\tau}_{L}}-m_{\tilde{\tau}_{R}}$, $m_{\tilde{\tau}_{L}}-m_{\tilde{\chi}_{1}^{\pm}}$, $m_{\tilde{l}_{L}}-m_{\tilde{l}_{R}}$ and $m_{\tilde{l}_{L}}-m_{\tilde{\chi}_{1}^{\pm}}$ planes, where $\tilde{l} = \tilde{e}, \tilde{\mu}$. The color coding is the same as in Figure \ref{fig6}. The diagonal lines represent the regions where the plotted masses are equal to each other.}
\label{fig:sleptons}
\end{figure}

\begin{figure}[ht!]
\centering
\subfigure[Signal 1]{\includegraphics[scale=1.5]{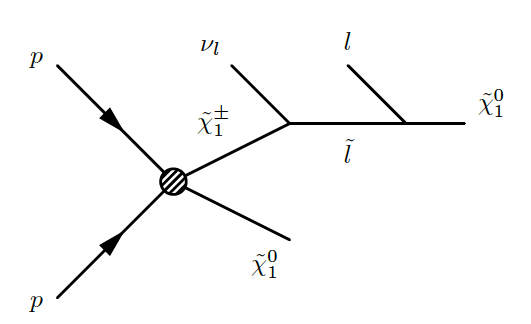}}
\subfigure[Signal 2]{\includegraphics[scale=1.5]{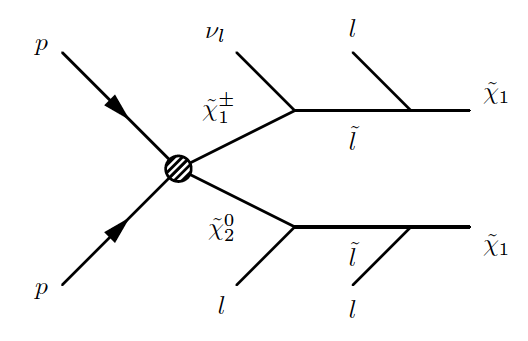}}
\caption{Chargino-Neutralino production processes involving the slepton in subsequent decays.}
\label{fig:signals}
\end{figure}

\begin{figure}[ht!]
\centering
\subfigure{\includegraphics[scale=0.4]{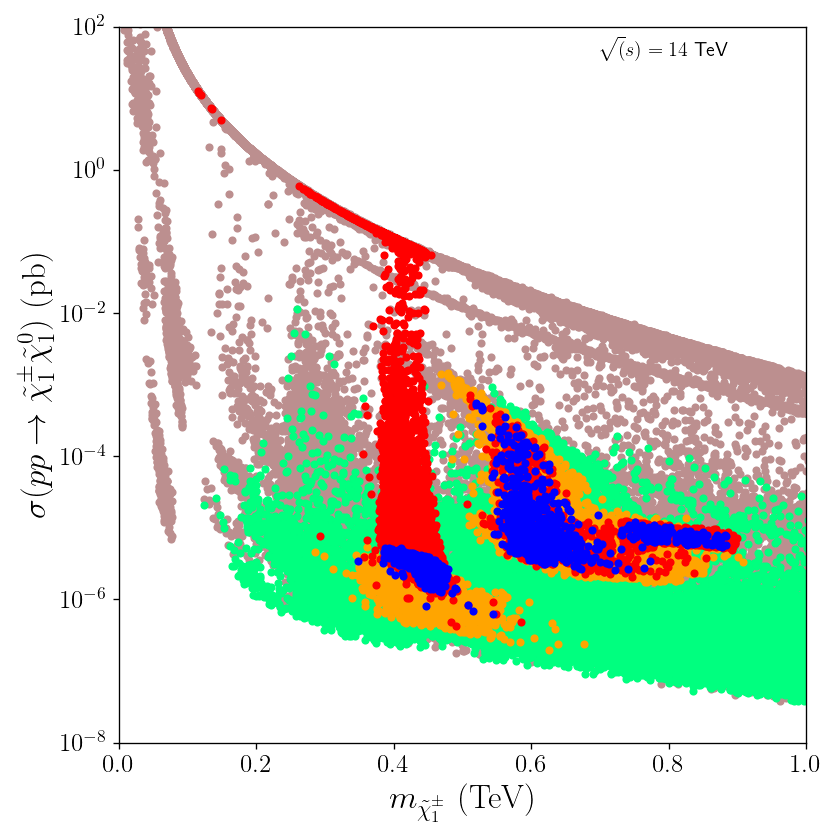}}%
\subfigure{\includegraphics[scale=0.4]{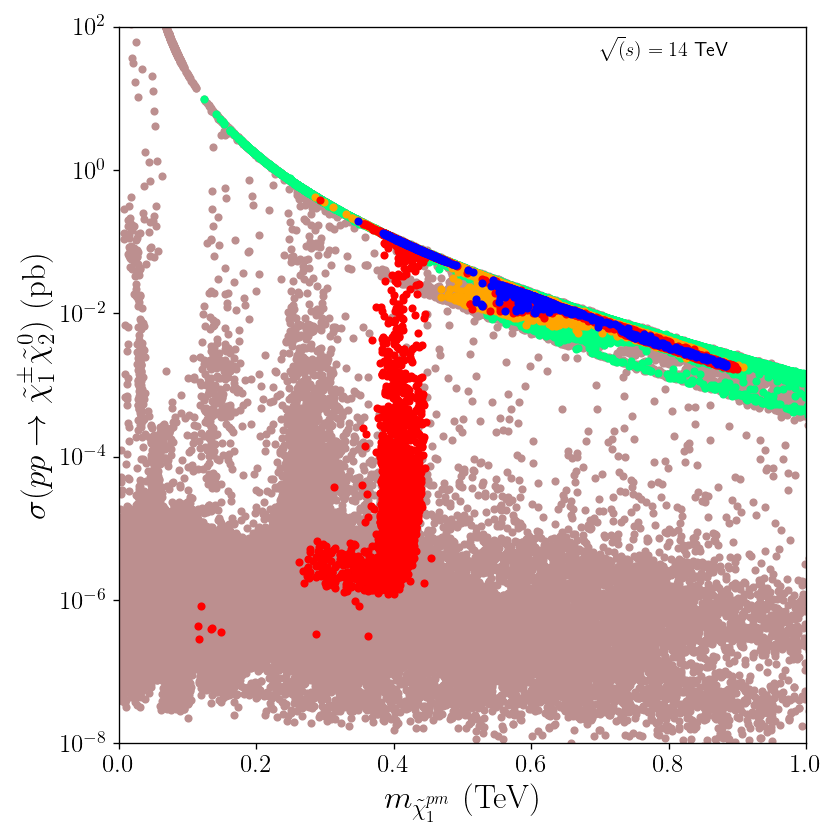}}
\caption{Chargino-Neutralino pair production at $\sqrt{s}=14$ TeV for Signal 1 (left) and Signal 2 (right). The color coding is the same as in Figure \ref{fig6}.}
\label{fig:CNprod}
\end{figure}

\begin{figure}[ht!]
\subfigure{\includegraphics[scale=0.4]{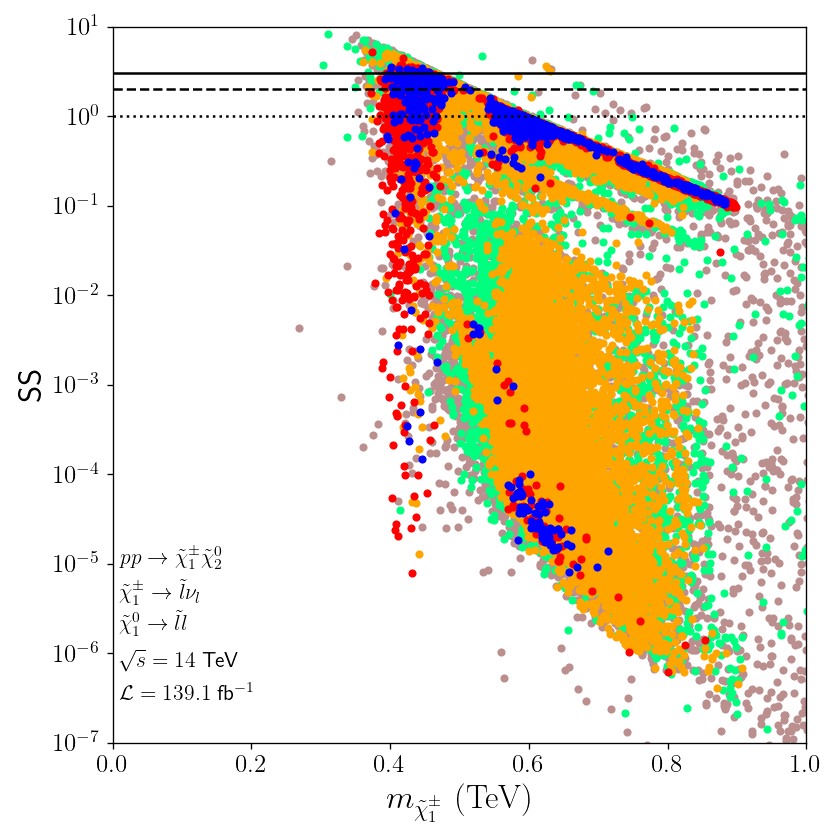}}%
\subfigure{\includegraphics[scale=0.4]{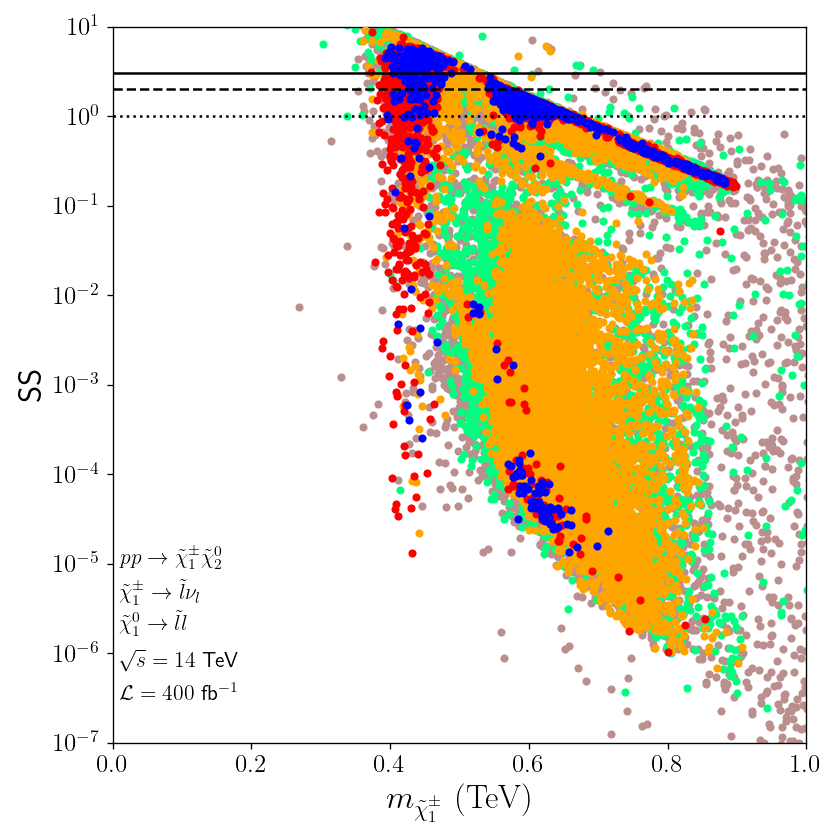}}
\subfigure{\includegraphics[scale=0.4]{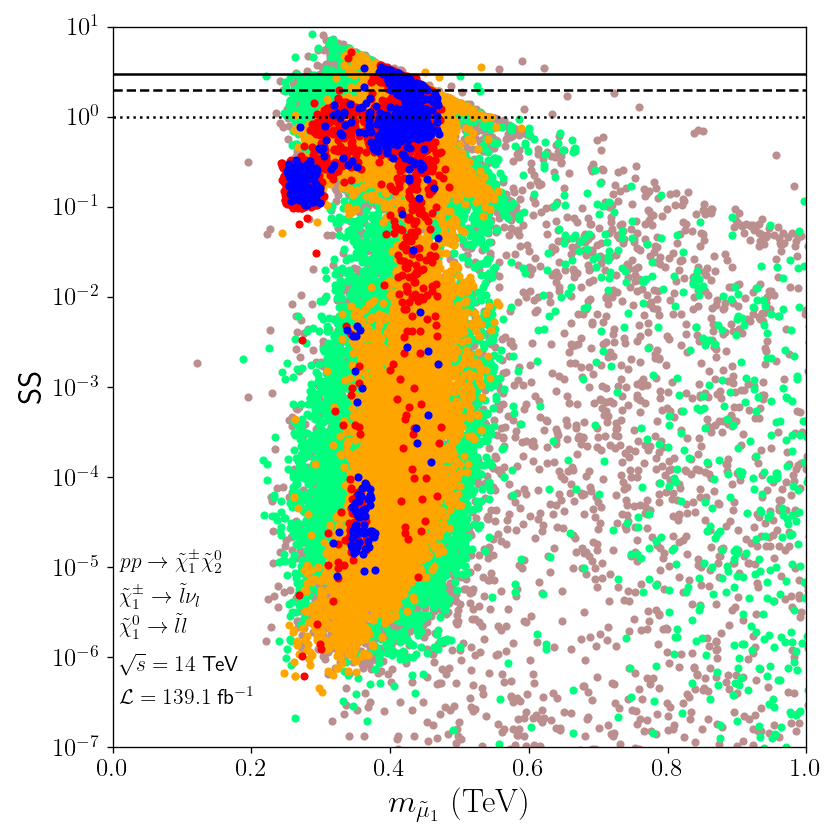}}%
\subfigure{\includegraphics[scale=0.4]{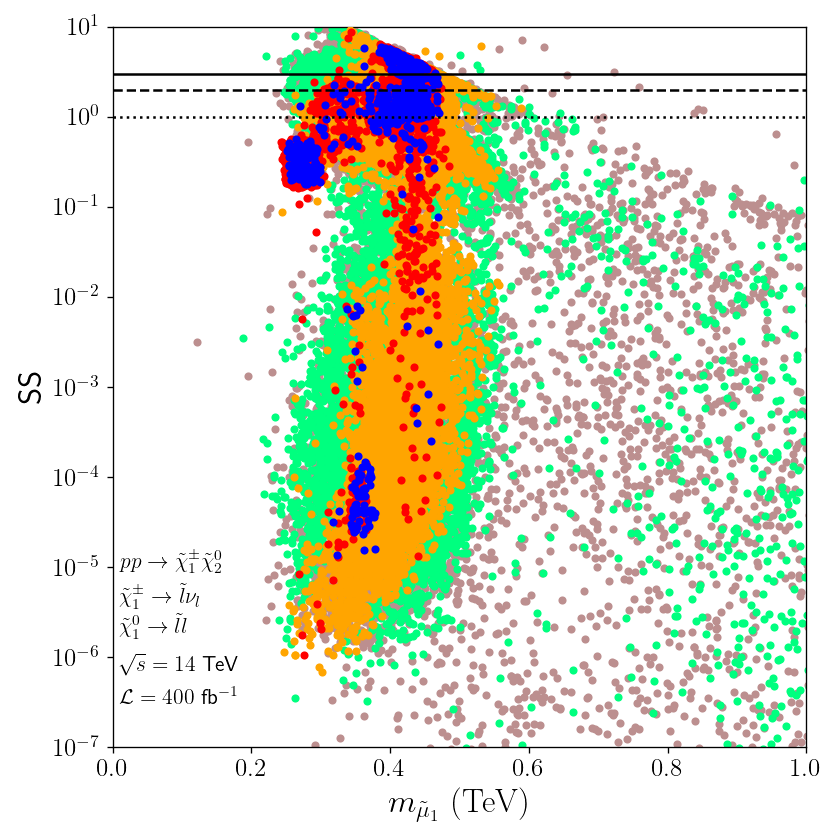}}
\caption{The expected probe on the chargino and slepton masses during the LHC-Run3 experiments. SS is calculated by Eq.(\ref{eq:SS}) and plotted in correlation with the chargino (top) and slepton (bottom) masses for 139.1 fb$^{-1}$ (left) and 400 fb$^{-1}$ (right) luminosities. The color coding is the same as Figure \ref{fig6}. The dotted, dashed and solid horizontal lines correspond to SS=1,2,3 respectively.}
\label{fig:SS}
\end{figure}

The SUSY spectra compatible with the muon $g-2$ resolution involve light sleptons, neutralinos and charginos, and the current LHC experiments are sensitive enough to probe the mass scales for these particles. The main impact in such regions of muon $g-2$ solutions comes from chargino-neutralino production at the collisions when the slepton is lighter than the chargino \cite{CMS:2017fdz,Sirunyan:2017zss,Aaboud:2018jiw}. In such processes, the chargino should be allowed to decay into a lighter slepton; however, the mass hierarchy between the chargino and slepton ($m_{\tilde{\chi}^{\pm}} \gtrsim m_{\tilde{l}}$) is not the only condition, the neutralino and chargino species also play a role in the signal processes. If the LSP neutralino is formed mostly by the Wino and/or Higgsino, then the spectrum typically involves $m_{\tilde{l}} > m_{\tilde{\chi}_{1}^{\pm}} \simeq m_{\tilde{\chi}_{1}^{0}}$; thus, the LSP neutralino should be mostly Bino-like to have sleptons lighter than charginos. With the Bino-like LSP neutralino the chargino can be either the Wino or Higgsino, each of which yields different impact on the sleptons. For instance, if the lighter chargino state is formed mostly by Higgsino, the chargino decays into sleptons through the Yukawa interactions. In this case, the collider experiments are not sensitive enough to probe the sleptons from the first two families due to the small Yukawa couplings. In addition, as we showed in Figure \ref{fig5}, the muon $g-2$ resolution in the MSSM requires $\mu \gtrsim 500$ GeV leading to heavy Higgsinos in the spectra which suppresses the chargino-neutralino production cross-section in the collisions \cite{Fiaschi:2018hgm}.

On the other hand, if the Wino forms the lighter chargino, its decays do not distinguish the sleptons since the interaction is the family universal $SU(2)_{L}$ gauge interaction. However, due to the $SU(2)_{L}$ symmetry, the lighter slepton state should mostly formed by the left-handed state such that the chargino can decay into the sleptons. Figure \ref{fig:sleptons} investigates the chirality of the sleptons with plots in the $m_{\tilde{\tau}_{L}}-m_{\tilde{\tau}_{R}}$, $m_{\tilde{\tau}_{L}}-m_{\tilde{\chi}_{1}^{\pm}}$, $m_{\tilde{l}_{L}}-m_{\tilde{l}_{R}}$ and $m_{\tilde{l}_{L}}-m_{\tilde{\chi}_{1}^{\pm}}$ planes, where $\tilde{l} = \tilde{e}, \tilde{\mu}$. The color coding is the same as in Figure \ref{fig6}. The diagonal lines represent the regions where the plotted masses are equal to each other. The $m_{\tilde{\tau}_{L}}-m_{\tilde{\tau}_{R}}$ plane shows that the left-handed staus weigh about 1 TeV and heavier, and the right-handed staus can be as light as about 300 GeV. This mass hierarchy results in that the lighter stau state is mostly right-handed in the muon $g-2$ region (orange). In addition, the left-handed stau is also heavier than the chargino in the same region, as shown in the $m_{\tilde{\tau}_{L}}-m_{\tilde{\chi}_{1}^{\pm}}$ plane. Consequently, the staus cannot take part in possible signal processes triggered by the chargino-nuetralino production in the regions, where the muon $g-2$ anomaly is resolved within the $1\sigma$ interval. 

In contrast to staus, the mass hierarchy between the left- and right-handed states is interchangeable  for the sleptons of the first two families. On the contrary, the muon $g-2$ solutions mostly yield left-handed states that are lighter than the right-handed ones as shown in the $m_{\tilde{l}_{L}}-m_{\tilde{l}_{R}}$ plane. The heaviest mass scales for these states are about 600 GeV for $m_{\tilde{\mu}_{L}}$ and 800 GeV for $m_{\tilde{\mu}_{R}}$. Even though it is possible to realize muon $g-2$ solutions with lighter states is light $m_{\tilde{\mu}_{L,R}} \gtrsim 200$ GeV, the compatible parameter space (orange) mostly prefers $m_{\tilde{\mu}_{L}} < m_{\tilde{\mu}_{R}}$. Besides, the $m_{\tilde{l}_{L}}-m_{\tilde{\chi}_{1}^{\pm}}$ plane shows that the left-handed slepton state is lighter than the chargino in most of the parameter space. 

As it turns out the muon $g-2$ solutions provide suitable topology for processes involving chargino decays into sleptons of the first two families, which is diagrammatically shown in Figure \ref{fig:signals}. Figure \ref{fig:CNprod} displays the chargino-neutralino production cross-section relevant to Signal 1 (left) and Signal 2 (right). The color coding is the same as in Figure \ref{fig6}. If the LSP neutralino is involved in the production process, it is possible to realize solutions with $\sigma(pp\rightarrow \tilde{\chi}_{1}^{\pm}\tilde{\chi}_{1}^{0}) \sim \mathcal{O}(10)$ pb as shown in the $\sigma(pp\rightarrow \tilde{\chi}_{1}^{\pm}\tilde{\chi}_{1}^{0})-m_{\tilde{\chi}_{1}^{\pm}}$ plane. However, such large cross-section values can be realized only in a very narrow region, and most of the muon $g-2$ solutions predict $\sigma(pp\rightarrow \tilde{\chi}_{1}^{\pm}\tilde{\chi}_{1}^{0}) \lesssim 10^{-2}$ pb. The $\sigma(pp\rightarrow \tilde{\chi}_{1}^{\pm}\tilde{\chi}_{2}^{0})-m_{\tilde{\chi}_{1}^{\pm}}$ plane shows that if $\tilde{\chi}_{2}^{0}$ is produced along with the chargino, the production cross-section cannot exceed 1 pb. However, most of the muon $g-2$ solutions are realized with $\sigma(pp\rightarrow \tilde{\chi}_{1}^{\pm}\tilde{\chi}_{2}^{0}) \gtrsim 10^{-2}$ pb even for $m_{\tilde{\chi}_{1}^{\pm}} \lesssim 700$ GeV. In this context, the processes assigned to Signal 2 are expected to provide the main impact in probing the muon $g-2$ region through leptonic decays of the chargino and neutralino, whose cross-section can be approximately calculated as:
\begin{equation}
\sigma({\rm Signal~2}) \approx \sigma(pp\rightarrow \tilde{\chi}_{1}^{\pm}\tilde{\chi}_{2}^{0})\times {\rm BR}(\tilde{\chi}_{1}^{\pm}\rightarrow \tilde{l}\nu_{l})\times {\rm BR}(\tilde{\chi}_{2}^{0}\rightarrow \tilde{l}l)\times \left[{\rm BR}(\tilde{l}\rightarrow \tilde{\chi}_{1}^{0}l)\right]^{2}~,
\label{eq:sig2}
\end{equation}
with an error of only about $0.7\%$ \cite{Altin:2019veq}.

We can quantify the Signal Strength (SS) by identifying the relevant SM background processes. Considering the Signal 2 processes the relevant SM background with multiple leptons in the final state is formed by the $\bar{t}t$ and $WW$, $WZ$ and $ZZ-$ pair production in the collisions. Comparing the top-quark pair production ($\sigma(pp\rightarrow \bar{t}t) \simeq 830$ pb \cite{Aad:2020tmz}) with $WW-$production ($\sigma(pp\rightarrow WW) \simeq 115.3$ pb \cite{CMS:2016vww}), $WZ-$production (($\sigma(pp\rightarrow WZ) \simeq 48.09$ pb \cite{Sirunyan:2019bez,Aaboud:2019gxl}) and $ZZ-$production ($\sigma(pp\rightarrow ZZ) \simeq 39.9$ pb \cite{Sirunyan:2017zjc}) the SM background is mainly formed by the top-quark pair production in which the top quarks subsequently decay into W-bosons. The signal strength can be quantified as 

\begin{equation}
SS=\dfrac{S}{\sqrt{S+B}}~,\hspace{0.5cm}
\setstretch{2}
\begin{array}{ll}
S& \equiv \sigma({\rm Signal~2})\times \mathcal{L} \\
B & \equiv \sigma({\rm Background})\times \mathcal{L}~,
\end{array}
\label{eq:SS}
\end{equation}
where $S$ and $B$ represent the event numbers of Signal 2 and Background processes respectively, as a function of the cross-sections ($\sigma({\rm Signal~2})$ and $\sigma({\rm Background})$) and the luminosity ($\mathcal{L}$). The SS can be translated into the confidence level (CL) by using the following correspondence \cite{Cranmer:2015nia}

\begin{equation}
\setstretch{1.5}
\begin{array}{rcl}
0 \leq & {\rm SS} & < 1\rightarrow {\rm hardly ~ probed} \\
1 \leq & {\rm SS} & < 2 \rightarrow {\rm probed~up~to~}68\%~{\rm CL}\\
2 \leq & {\rm SS} & < 3 \rightarrow {\rm probed~up~to~}95\%~{\rm CL} \\
3 \leq & {\rm SS} &  \hspace{0.7cm}\rightarrow {\rm Discovery.}
\end{array}
\end{equation}

We show our calculations for SS in correlation with the chargino (top) and slepton (bottom) masses for luminosities 139.1 fb$^{-1}$ (left) and 400 fb$^{-1}$ (right) in Figure \ref{fig:SS}. The color coding is the same as in Figure \ref{fig6}. The dotted, dashed and solid horizontal lines correspond to SS=1,2,3 respectively. The details about the production and the decay processes along with the center of mass energy ($\sqrt{s}$) and luminosities ($\mathcal{L}$) are given in the botom-left of each plot. The current luminosity has reached 139.1 fb$^{-1}$, and the targeted luminosity is 400 fb$^{-1}$ at the end of the Run3 experiments at LHC. A direct impact from Signal 2 processes is observed on the chargino mass as the top-left plane shows that the current experiments are able to probe the chargino mass up to about 600 GeV at the $68\%$ CL and 500 GeV at the $95\%$ CL in the muon $g-2$ region. The Run3 experiments can probe the chargino mass up to about 700 GeV at the $68\%$ CL and 600 GeV at the $95\%$ CL, as seen from the top-right panel which plots the chargino mass and the SS values for 400 fb$^{-1}$. Depending on the slepton mass and the mixing of the right-handed slepton in the lighter state can suppress the SS of Signal 2 so that it is still possible to realize the muon $g-2$ solutions within a 600-700 GeV chargino mass range, which can escape from the probe. Indeed, these solutions are realized with large $\mu\tan\beta$, which enhances the SUSY contributions to muon $g-2$ and the mixing between $\tilde{\mu}_{L}-\tilde{\mu}_{R}$.

The bottom planes of Figure \ref{fig:SS} show the SS values in correlation with the slepton masses, and our results show that the impact on the slepton mass is rather indirect. The sensitivity in the experiments can provide a bound on the sleptons given by $m_{\tilde{l}} \gtrsim 450$ GeV. This mass bound seems not to change too much during the Run3 experiments, but these experiments will be able to extend the muon $g-2$ region to be probed as there are more solutions with ${\rm SS} \geq 1$ in the bottom right panel compared to the bottom-left one.

\section{Dark Matter Implications}
\label{sec:DM}

\begin{figure}[htb!]
\centering
\subfigure{\includegraphics[scale=0.4]{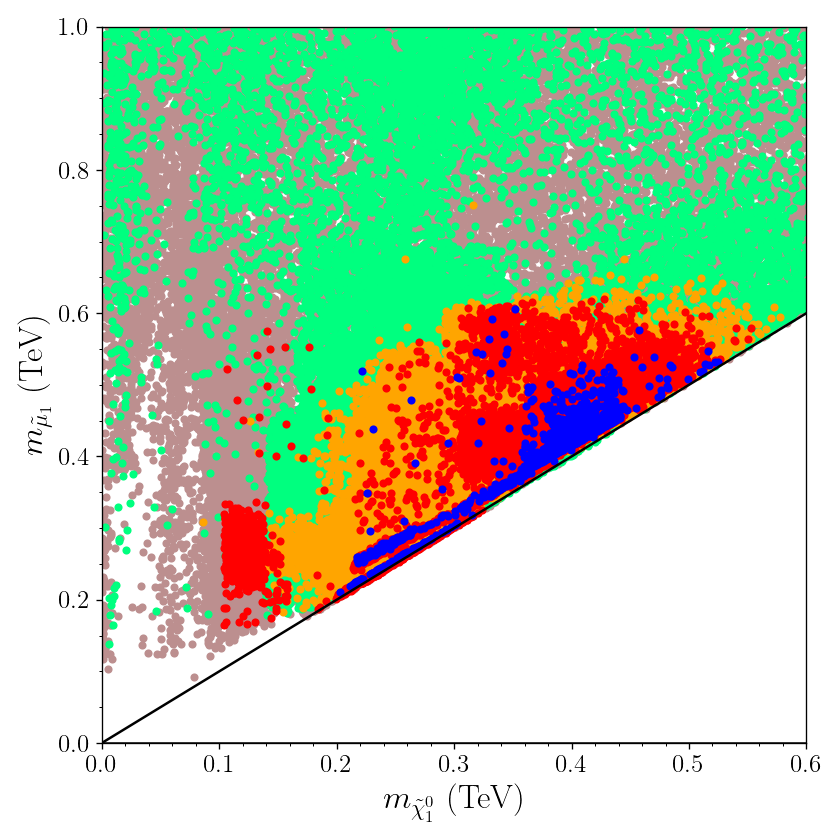}}%
\subfigure{\includegraphics[scale=0.4]{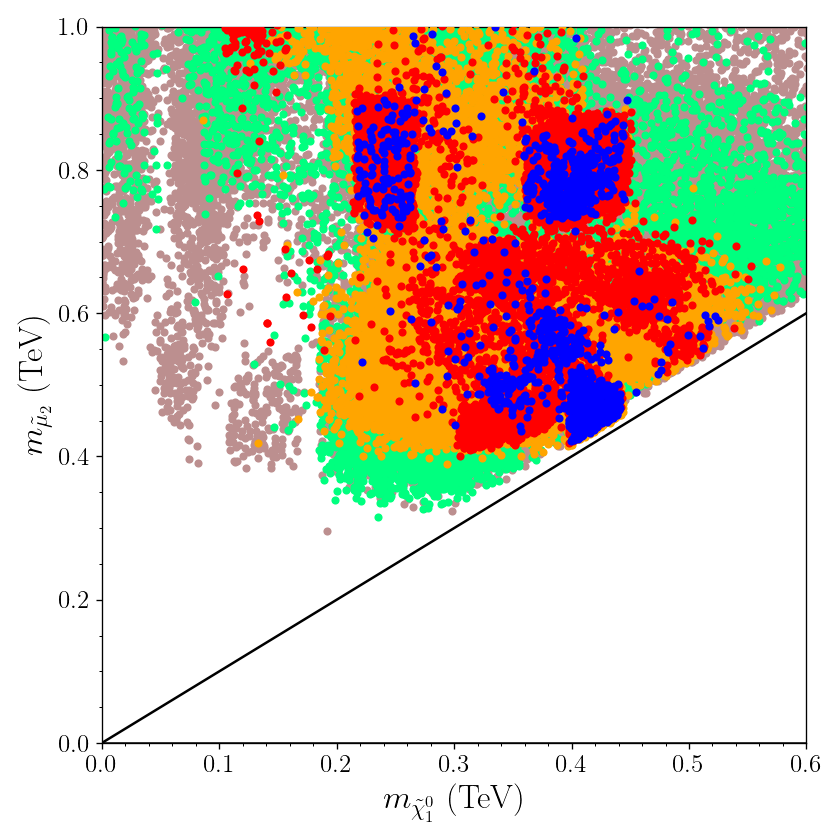}}
\subfigure{\includegraphics[scale=0.4]{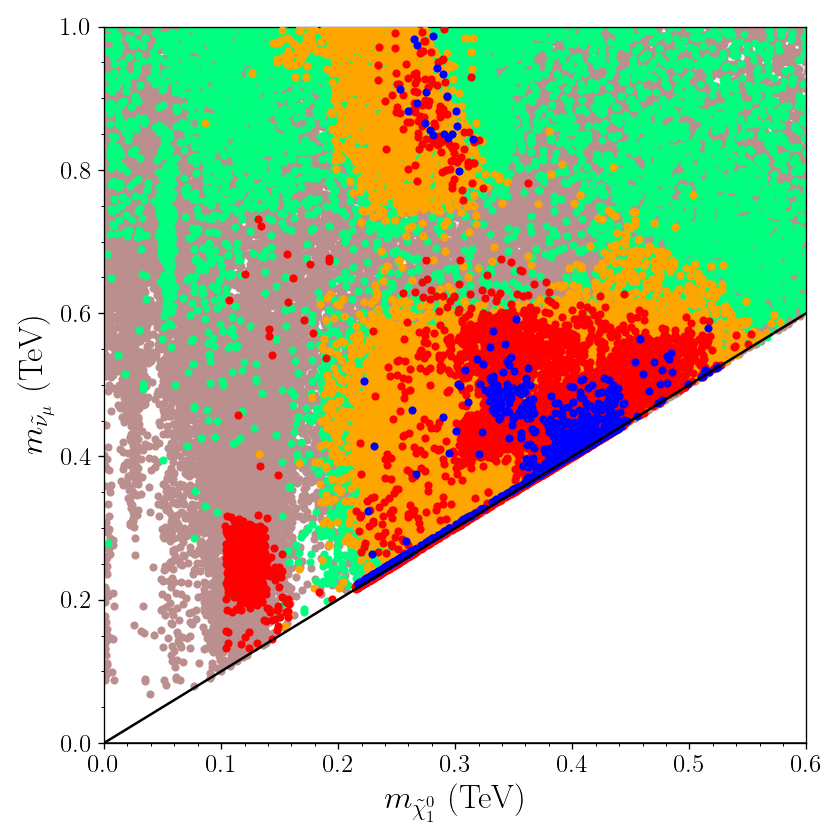}}%
\subfigure{\includegraphics[scale=0.4]{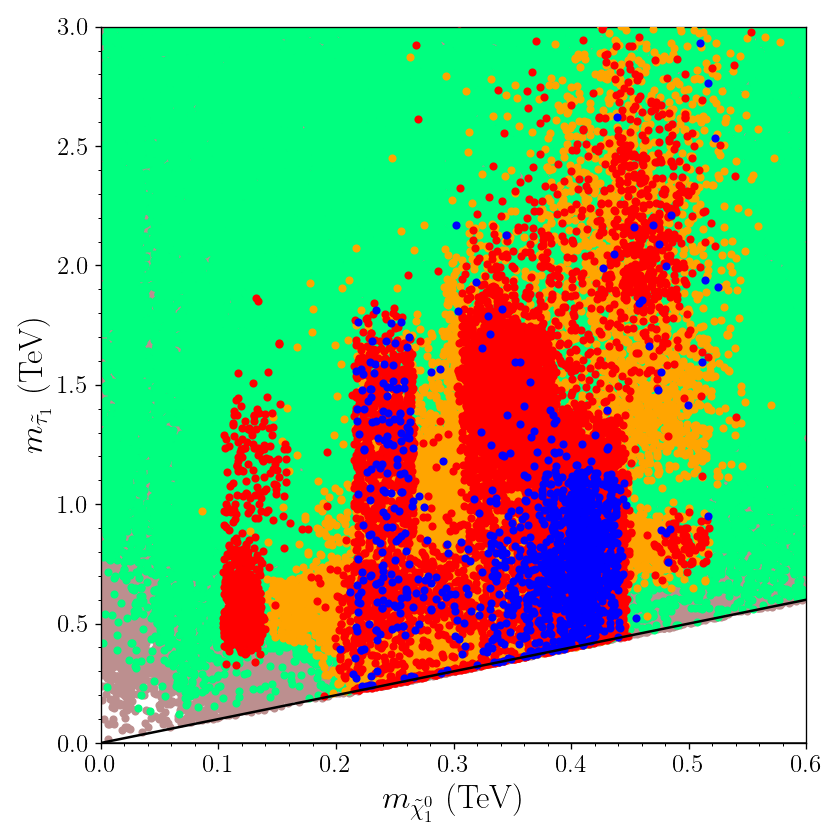}}
\caption{Smuon and sneutrino masses with plots in the $m_{\tilde{\mu}_{1}}-m_{\tilde{\chi}_{1}^{0}}$, $m_{\tilde{\mu}_{2}}-m_{\tilde{\chi}_{1}^{0}}$, $m_{\tilde{\nu}_{\mu}}-m_{\tilde{\chi}_{1}^{0}}$ and $m_{\tilde{\tau}_{1}}-m_{\tilde{\chi}_{1}^{0}}$ planes. The color coding is the same as in Figure \ref{fig6}. The diagonal lines show the mass degeneracy between the plotted particles in each plane.}
\label{fig7}
\end{figure}

\begin{figure}[htb!]
\centering
\subfigure{\includegraphics[scale=0.4]{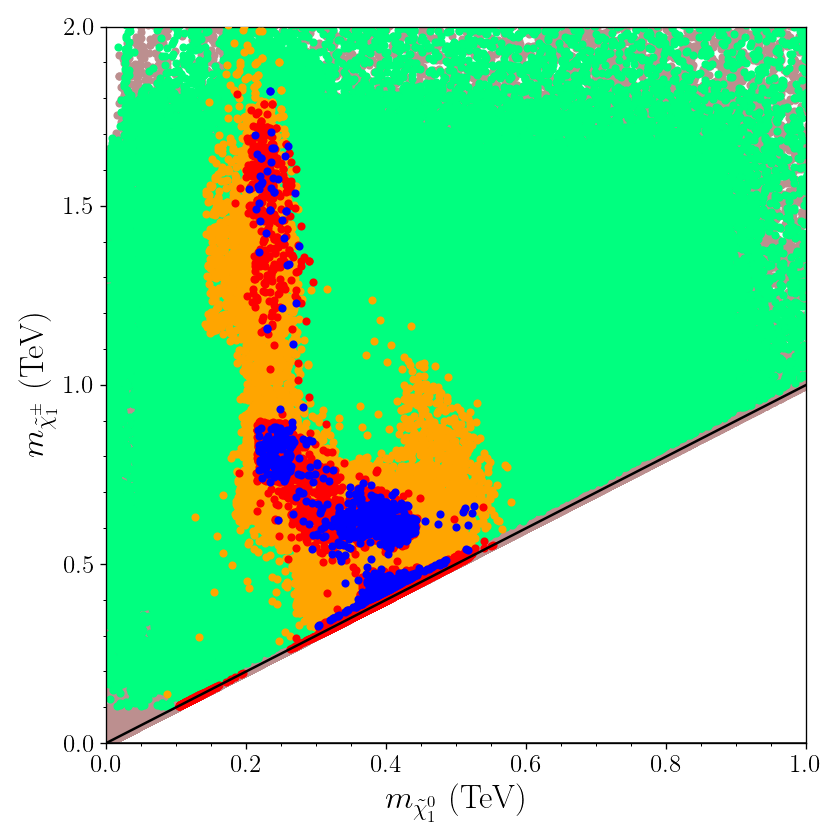}}%
\subfigure{\includegraphics[scale=0.4]{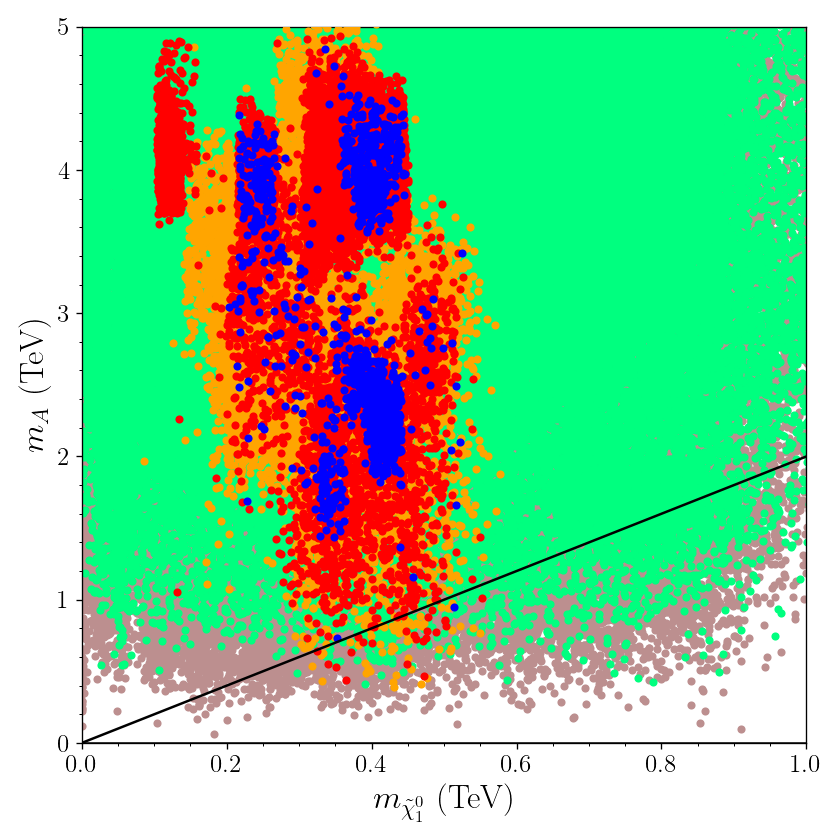}}
\caption{Plots in the $m_{\tilde{\chi}_{1}^{\pm}}-m_{\tilde{\chi}_{1}^{0}}$ and $m_{A}-m_{\tilde{\chi}_{1}^{0}}$ planes. The color coding is the same as in Figure \ref{fig7}.}
\label{fig8}
\end{figure}

\begin{figure}[htb!]
\centering
\subfigure{\includegraphics[scale=0.4]{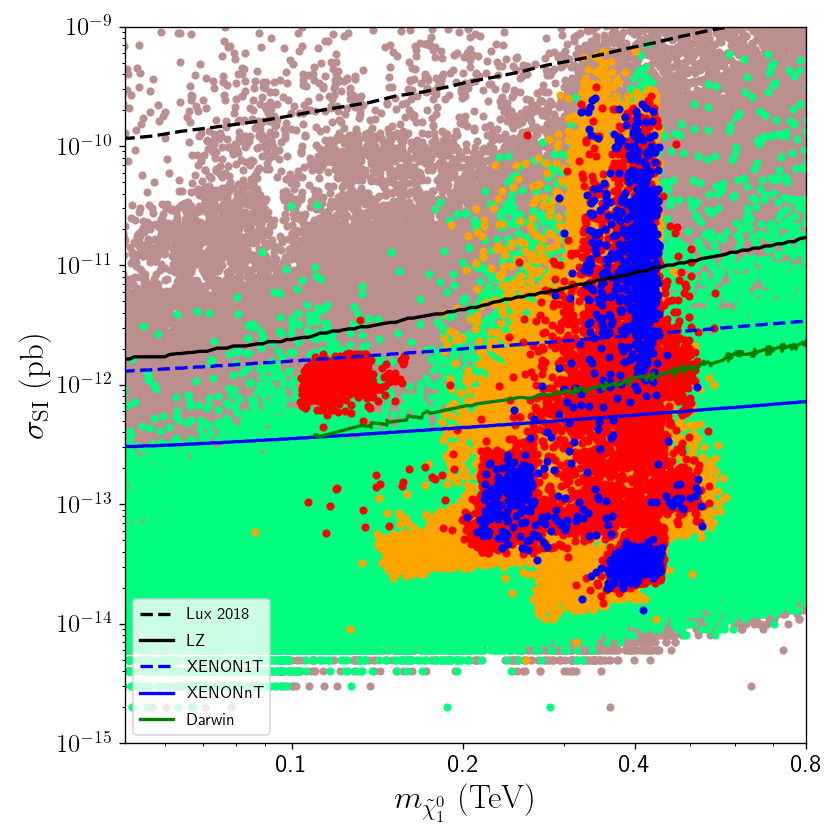}}%
\subfigure{\includegraphics[scale=0.4]{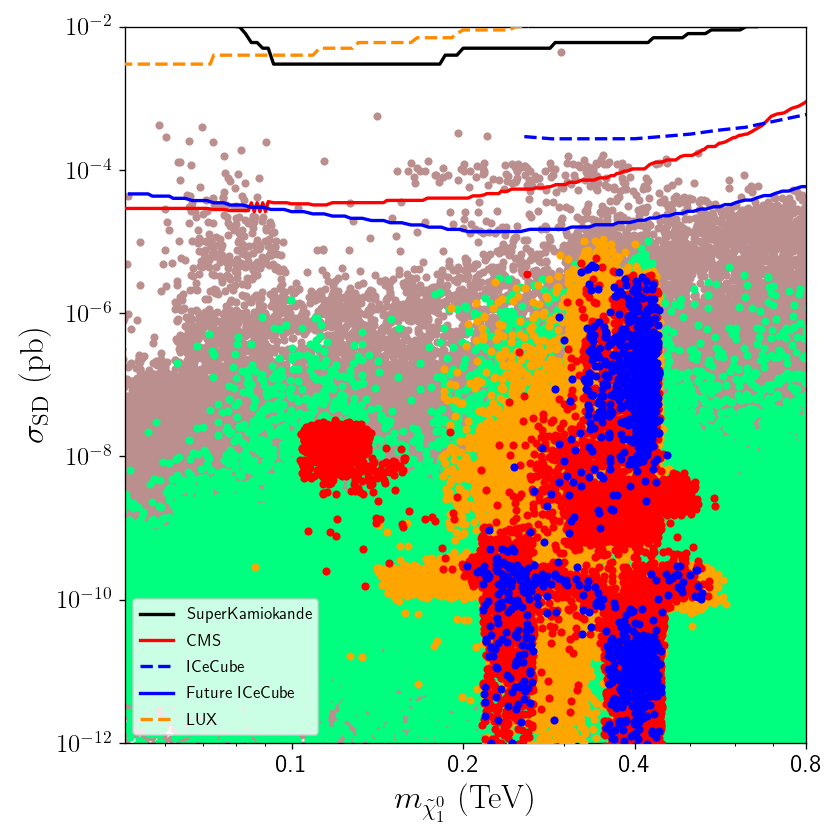}}
\caption{Plots for the spin-independent (left) and spin-dependent (right) scattering cross-sections of the dark matter in the $\sigma_{{\rm SI}}$ and $\sigma_{{\rm SD}}-m_{\tilde{\chi}_{1}^{0}}$ planes. The color coding is the same as in Figure \ref{fig6}.}
\label{fig9}
\end{figure}

As we discussed in the previous section, a resolution of the muon $g-2$ discrepancy within the $1\sigma$ interval requires the Bino to be lighter than about 600 GeV, which allows it to significantly participate in forming the LSP neutralino. In this case the current measurements on the relic density of dark matter can provide a challenge since Bino usually has a large relic density \cite{Ibe:2013oha}. Realizing the correct (or as we allowed, a lower) relic density of LSP neutralino requires some suitable coannihilation scenarios in which NLSP is close in mass to the LSP. Indeed, the Bino-like LSP solutions need multiple coannihilation scenarios to satisfy the relic density condition. We discuss possible coannihilation scenarios in Figure \ref{fig7} by plotting the smuon, sneutrino and stau masses versus the LSP neutralino mass. The color coding is the same as in Figure \ref{fig6}. The diagonal lines show the mass degeneracy between the plotted particles in each plane. The $m_{\tilde{\mu}_{1}}-m_{\tilde{\chi}_{1}^{0}}$ plane shows that the lightest smuon happens to be the NLSP in much of the parameter space compatible with the muon $g-2$ constraint within $1\sigma$, and the LSP neutralino can co-annihilate with the NLSP smuon when its mass lies between about 100 and 500 GeV (red and blue points). On the other hand, if we require the correct relic density, then a consistent smuon-nuetralino coannihilation scenario can be observed for $m_{\tilde{\mu}_{1}} \gtrsim 240$ GeV (blue points). The second smuon cannot be lighter than about 400 GeV if it is close to the LSP neutralino in the mass spectrum as shown in the $m_{\tilde{\mu}_{2}}-m_{\tilde{\chi}_{1}^{0}}$ plane. However, it still helps in decreasing the LSP neutralino relic abundance through the co-annihilation mechanisms around its mass scale. In addition, the sneutrinos can also take part in the coannihiation scenarios. With coannihilation, the relic abundance of the LSP neutralino is significantly decreased such that $\Omega h^{2}_{LSP} < 0.114$, as seen in the red region for $m_{\tilde{\nu}_{\mu}} \in [100-200]$ GeV in the $m_{\tilde{\nu}_{\mu}}-m_{\tilde{\chi}_{1}^{0}}$ plane. Insisting on the correct relic density excludes this region, and the snuetrino-neutralino coannihilation scenario can be realized for $m_{\tilde{\nu}_{\mu}} \gtrsim 250$ GeV. The $m_{\tilde{\tau}_{1}}-m_{\tilde{\chi}_{1}^{0}}$ plane shows that there are also solutions for the stau-neutralino coannihilation scenario for $m_{\tilde{\tau}_{1}}$ in the 200-450 GeV range.

In addition to the sleptons, we identify solutions for chargino-neutralino coannihilation scenario as shown in the $m_{\tilde{\chi}_{1}^{\pm}}-m_{\tilde{\chi}_{1}^{0}}$ plane of Figure \ref{fig8}. The color coding is the same as in Figure \ref{fig7}. As we discussed in the section on the mass spectrum, a resolution of the muon $g-2$ anomaly also favors light Winos, even though it is still possible to have heavy $SU(2)_{L}$ gauginos. The solutions around the diagonal lines indicate the charginos and neutralinos that are degenerate in mass. Although the mass degeneracy between the chargino and LSP neutralino is realized if the LSP neutralino is composed of the Wino or Higgsino in the class of SUSY GUTs with universal gaugino masses, it is possible to realize nearly mass degenerate Bino-like LSP neutralino and chargino in our scans since $M_{1}$ and $M_{2}$ are independent parameters. The solutions far above the diagonal line correspond to the region of muon $g-2$ where the Wino happens to be heavy, and the LSP neutralino is mostly a Bino. Figure \ref{fig8} also shows that the $A-$resonance solutions are possible in the framework of our model. The solutions around the diagonal line have $m_{A} \simeq 2m_{\tilde{\chi}_{1}^{0}}$, and they can be identified for $ 360 \gtrsim m_{\tilde{\chi}_{1}^{0}} \gtrsim 600$ GeV. These solutions can be probed up to $m_{A} \sim 800$ GeV through hadronic decays of $A-$boson \cite{Aad:2019zwb, Sirunyan:2019arl}, and $m_{A} \sim 1$ TeV through leptonic decays \cite{Aad:2020zxo,Aaboud:2018gjj}.

If the Bino-like LSP is realized with the correct (or lower) relic density, then it can also be consistent with the latest constraints from the direct detection of dark matter experiments, since it yields low cross-sections in dark matter scattering.  We display our results for the spin-independent and spin-dependent scattering cross-section in Figure \ref{fig9} with plots in the $\sigma_{{\rm SI}}$ and $\sigma_{{\rm SD}}-m_{\tilde{\chi}_{1}^{0}}$ planes. The color coding is the same as in Figure \ref{fig6}. We also display the current and projected exclusion curves from several experiments such as LUX and LZ \cite{Akerib:2018lyp}, XENON \cite{Aprile:2020vtw} and DARWIN \cite{Aalbers:2016jon} for the spin-independent scattering cross-sections. The experimental curves and the corresponding experiments are given in the legend. As seen from the $\sigma_{{\rm SI}}-m_{\tilde{\chi}_{1}^{0}}$ plane, the muon $g-2$ solutions (orange, red and blue) predict cross-sections lower than the current LUX bounds, and some of them lie above the projected reach of the LZ experiment. These solutions are expected to be tested after the ongoing upgrades of the LUX experiment \cite{Akerib:2020com,Akerib:2020ewf}. In addition, DARWIN has already provided more sensitive results, which excludes the solutions leading to a dark matter scattering cross-section greater than about $10^{-12}$ pb. Some of these solutions still lie between the current and projected results from the XENON experiment, and they will potentially be tested soon by the XENON collaboration. The remaining solutions with Bino-like dark matter will require further upgrades in direct detection experiments in order to be tested.

We also display our results for the spin-dependent cross-sections in the right panel of Figure \ref{fig9} versus some experimental results from SuperKamiokande \cite{Tanaka:2011uf}, collider experiments \cite{Khachatryan:2014rra}, IceCube \cite{Abbasi:2009uz} and LUX \cite{Akerib:2016lao} experiments. Even though they are not as sensitive as the experimental results for the spin-independent scattering, some solutions can be tested after some further upgrade in IceCube.

Before concluding, we exemplify our findings with benchmark points given in two tables. Table \ref{tab1} represents six benchmark points which are chosen to be consistent with all the applied constraints including the Planck dark matter bounds. The masses are given in GeV and the cross-sections in pb. The benchmark points listed in Table \ref{tab1} lead to a Bino LSP, and the spectra involve a variety of light supersymmetric particles which can be nearly degenerate in mass with the LSP neutralino. Point 1 depicts a solution which is in accord with the Planck bound through the slepton-neutralino coannihilation scenario involving the lighter slepton states from the first two families. Point 2 also exemplifies the solutions for the slepton-neutralino coannihilation scenario with two sleptons involved in the coannihilation processes. Point 3 represents a solution for the stau-neutralino coannihilation scenario, while the first two-family sleptons are slightly heavier for such solutions. Point 4 depicts the chargino-neutralino coannihilation scenario. Point 5 also exemplifies solutions of the chargino-neutralino scenario, but it also provides an example for $A-$resonance solutions. All these solutions usually yield a relatively low spin-independent cross-section for dark matter scattering, which needs further upgrades in the direct detection experiments to be tested. We display in Point 6 the solutions that predict a relatively large spin-independent scattering cross-section of dark matter which can be tested in the near future. The $\mu-$term is about 500 GeV for such solutions, and the spectrum involves light Higgsinos.

\begin{table}[h!]
\centering
\setstretch{1.2}
\scalebox{0.9}{
\begin{tabular}{|c|cccccc|}
\hline  & Point 1 & Point 2 & Point 3 & Point 4 & Point 5 & Point 6 \\ \hline
$m_{0_{1,2}}$ & 325 & 112.3 & 160.3 & 499.5 & 444.7 & 120.7 \\
$m_{0_{3}}$ & 1989 & 2166 & 2000 & 3025 & 2472 & 1893 \\
$M_{1}$ & 854.8 & 1010 & 817.8 & 885 & 1073 & 977.9 \\
$M_{2}$ & 483.4 & 759.4 & 721.1 & 523 & 514 & 745.6 \\
$M_{3}$ & 2139 & 2079 & 1764 & 2691 & -3891 & 1956 \\
$A_{0}/m_{0_{3}}$ & -3.0 & -2.0 & -2.2 & -2.7 & -1.4 & -1.3 \\
$\tan\beta$ & 20.2 & 44.9 & 43.5 & 43.3 & 44.4 & 47.8 \\ \hline
$\mu$ & 4508 & 1861 & 1537 & 5039 & 4080 & {\color{red}468.9} \\
$\Delta a_{\mu}\times 10^{10}$ & 24.6 & 25.6 & 28.9 & 26.1 & 24.3 & 22.3 \\ \hline
$m_{h}$ & 125.6 & 124.4 & 124.4 & 125.6 & 123.1 & 123.4 \\
$m_{H}$ & 3963 & 2540 & 2255 & 4148 & 946.8 & 2062 \\
$m_{A}$ & 3964 & 2540 & 2255 & 4148 & {\color{red}946.8} & 2062 \\
$m_{H^{\pm}}$ & 3967 & 2542 & 2258 & 4150 & 951.1 & 2065 \\ \hline
$m_{\tilde{\chi}_{1}^{0}}$,$m_{\tilde{\chi}_{2}^{0}}$ & {\color{red}365.2}, 387 & {\color{red}433.1}, 617.7 & {\color{red}348.8}, 587.4 & {\color{red}380.6}, {\color{red}419.9} & {\color{red}513.7}, 540.5 & {\color{red}411.2}, {\color{red}562.8} \\
$m_{\tilde{\chi}_{3}^{0}}$,$m_{\tilde{\chi}_{4}^{0}}$ & 4495, 4495 & 1909, 1911 & 1576, 1578 & 5028, 5029 & 4160, 4160 & 639.2, 682 \\
$m_{\tilde{\chi}_{1}^{\pm}}$,$m_{\tilde{\chi}_{2}^{\pm}}$ & 387.2, 4496 & 617.9, 1912 & 587.6, 1579 & {\color{red}420.1}, 5029 & 540.7, 4160 & {\color{red}561.5}, 680.8 \\
$m_{\tilde{g}}$ & 4470 & 4344 & 3724 & 5560 & 7841 & 4108 \\ \hline
$m_{\tilde{u}_{1}}$,$m_{\tilde{u}_{2}}$ & 3843, 3849 & 3719, 3740 & 3202, 3224 & 4732, 4759 & 6634, 6637 & 3518, 3545 \\
$m_{\tilde{t}_{1}}$,$m_{\tilde{t}_{2}}$ & 2272, 3323 & 2755, 3246 & 2218, 2740 & 3671, 4346 & 6190, 6274 & 2778, 3104 \\  \hline
$m_{\tilde{d}_{1}}$,$m_{\tilde{d}_{2}}$ & 3840, 3844 & 3726, 3741 & 3206, 3225 & 4760, 4767 & 6634, 6636 & 3530, 3546 \\
$m_{\tilde{b}_{1}}$,$m_{\tilde{b}_{2}}$ & 3298, 4096 & 3218, 3582 & 2709, 3077 & 4321, 4873 & 6230, 6324 & 3076, 3287 \\  \hline
$m_{\tilde{\nu}_{e}}$,$m_{\tilde{\nu}_{\mu}}$ & 467.9, 470.6 & 434.4, 442 & 444.1, 450.8 & 385.6, 400.1 & 520.7, 522.4 & 414.4, 420.3 \\
$m_{\tilde{l}_{1}}$,$m_{\tilde{l}_{2}}$ & {\color{red}378.9}, 475.7 & {\color{red}435.5}, {\color{red}460.7} & 370, 452.3 & 392.4, 812.6 & 522.8, 596.9 & 422, 482.2 \\
$m_{\tilde{\tau}_{1}}$,$m_{\tilde{\tau}_{2}}$ & 1512, 1809 & 502.4, 1608 & {\color{red}351.2}, 1484 & 1128, 2228 & 1937, 2251 & 551.8, 1418 \\  \hline
$\sigma_{{\rm SI}}$ & $ 7.9 \times 10^{-14} $ & $ 1.28 \times 10^{-12} $ & $ 2.02 \times 10^{-12} $ & $ 2.9 \times 10^{-14} $ & $ 9.56 \times 10^{-13} $ & $ 2.25 \times 10^{-10} $ \\
$\sigma_{{\rm SD}}$ & $ 7.08 \times 10^{-12} $ & $ 8.25 \times 10^{-9} $ & $ 1.85 \times 10^{-8} $ & $ 3.09 \times 10^{-13} $ & $ 2.57 \times 10^{-10} $ & $ 3.22 \times 10^{-6} $ \\
$\Omega h^{2}$ & 0.115 & 0.117 & 0.121 & 0.118 & 0.115 & 0.12 \\ \hline
\end{tabular}}
\caption{Benchmark points that exemplify our findings. The points are chosen to be consistent with all the applied constraints including the Planck bound on dark matter. The masses are given in GeV and the cross-sections in pb. Point 1 depicts a solution  which is in accord with the Planck bound through the slepton-neutralino coannihilation scenario involving the lighter slepton states from the first two families. Point 2 also exemplifies the solutions for the slepton-neutralino coannihilation scenario with two sleptons involved in the coannihilation processes. Point 3 represents a solution for the stau-neutralino coannihilation scenario, and the first two-family sleptons are slightly heavier for such solutions. Point 4 depicts the chargino-neutralino coannihilation scenario. Point 5 also exemplifies solutions of the chargino-neutralino coannihilation, but it also provides an example for $A-$resonance solutions. Point 6 represents solutions that predict a large spin-independent dark matter scattering cross-section, which can be tested in the near future.}
\label{tab1}
\end{table}

We also present four benchmark points in Table \ref{tab2} that summarize how the muon $g-2$ compatible region can be tested during the LHC-Run3 experiments. These points are chosen to be consistent with the applied constraints including the Planck bound on dark matter abundance. The subscripts on the SS symbols indicate the luminosity values. Point 1 depicts a solution with ${\rm SS} \simeq 1.04$. Such solutions can be probed in the current collider experiments or in the early runs of the Run3 experiments at LHC. The relevant SUSY particles weigh about 450 GeV, the chargino-neutralino production cross-section is about 0.072 pb, and although the signal strength for such solutions will be improved during the Run3 experiments, these solutions can be probed at the $95\%$ CL or higher in the later stages of the LHC experiments. Point 2 represents a solution for lighter sleptons ($\sim 250$ GeV) with the chargino as heavy as about 900 GeV. Despite the light slepton masses, such solutions can escape detection due to the heavy charginos. Point 3 displays a solution that remains viable even after Run3, despite improvements in the signal strength. Finally, point 4 exemplifies a solution which potentially will be probed at the end of Run3 at $68\%$ CL. Table \ref{tab2} also shows the branching ratios for the decay modes $\tilde{\chi}_{1}^{\pm}\rightarrow W^{\pm}\tilde{\chi}_{1}^{0}$, $\tilde{\chi}_{2}^{0}\rightarrow Z\tilde{\chi}_{1}^{0}$ and $\tilde{\chi}_{2}^{0}\rightarrow h\tilde{\chi}_{1}^{0}$. All these decay channels have branching ratios less than about $0.3\%$, and thus the analyses for processes involving these decays of $\tilde{\chi}_{1}^{\pm}$ and $\tilde{\chi}_{2}^{0}$ \cite{Aaboud:2018ngk,Robichaud-Veronneau:2016pzs} are not applicable for the solutions exemplified in Table \ref{tab2}. All these solutions reside in the slepton-neutralino coannihilation region in which the lighter chargino is mostly Wino-like.


\begin{table}[h!]
\centering
\setstretch{1.2}
\begin{tabular}{|c|cccc|}
\hline  & Point 1 & Point 2 & Point 3 & Point 4 \\ \hline
$m_{0_{1,2}}$ & 474.4 & 110.3 & 115.9 & 91 \\
$m_{0_{3}}$ & 2824 & 2510 & 1933 & 2028 \\
$M_{1}$ & 965.2 & 557.9 & 956.9 & 1008 \\
$M_{2}$ & 551.1 & 1062 & 687.3 & 816.6 \\
$M_{3}$ & 2837 & 1435 & 1969 & 2001 \\
$A_{0}/m_{0_{3}}$ & -2.9 & -2.5 & -1.2 & -1.7 \\
$\tan\beta$ & 44.4 & 40.7 & 47.3 & 44 \\ \hline
$m_{\tilde{\chi}_{1}^{0}}$,$m_{\tilde{\chi}_{2}^{0}}$ & {\color{red} 415.6}, 442.5 & {\color{red} 240}, 881.7 & {\color{red} 401.8}, 528.9 & {\color{red} 431.1}, 661.7 \\
$m_{\tilde{\chi}_{1}^{\pm}}$,$m_{\tilde{\chi}_{2}^{\pm}}$ & 442.7, 5057 & 881.9, 3068 & 527.7, 675 & 661.9, 1398 \\
$m_{\tilde{l}_{1}}$,$m_{\tilde{l}_{2}}$ & {\color{red} 435.8}, 754.8 & {\color{red} 256.8}, 885.7 & {\color{red} 407.2}, 416.2 & {\color{red} 440}, 524 \\  \hline
$\sigma(pp\rightarrow \tilde{\chi}_{1}^{\pm}\tilde{\chi}_{2}^{0})$ & 0.072 & 0.002 & 0.027 & 0.011 \\
${\rm BR}(\tilde{\chi}_{1}^{\pm}\rightarrow \tilde{l}\nu)$ & 0.468 & 0.99 & 0.457 & 0.968 \\
${\rm BR}(\tilde{\chi}_{2}^{0}\rightarrow \tilde{l}l)$ & 0.468 & 0.99 & 0.457 & 0.968 \\
${\rm BR}(\tilde{l}\rightarrow \tilde{\chi}_{1}^{0}l)$ & 1 & 1 & 1 & 1 \\ \hline
$\sigma({\rm Signal~2})$ & 0.008 & 0.001 & 0.003 & 0.005 \\
${\rm SS}_{139.1}$ & 1.04 & 0.12 & 0.44 & 0.67 \\
${\rm SS}_{400}$ & 1.76 & 0.2 & 0.75 & 1.14 \\ \hline 
${\rm BR}(\tilde{\chi}_{1}^{\pm}\rightarrow W^{\pm}\tilde{\chi}_{1}^{0})$ & 0.0 & 0.0 & 0.028 & 0.001 \\
${\rm BR}(\tilde{\chi}_{2}^{0}\rightarrow Z\tilde{\chi}_{1}^{0})$ & 0.0 & 0.0 & 0.001 & 0.0 \\
${\rm BR}(\tilde{\chi}_{2}^{0}\rightarrow h\tilde{\chi}_{1}^{0})$ & 0.0 & 0.0 & 0.015 & 0.001 \\ \hline
\end{tabular}
\caption{Benchmark points summarizing probe of the muon $g-2$ compatible region during the Run3 experiments at LHC. All points are chosen to be consistent with the applied constraints including the Planck bound on dark matter abundance. The masses are given in GeV and the cross-sections in pb. The subscripts on the SS symbols indicate the luminosity values. Point 1 depicts a solution which is about to be probed in the collider experiments. Point 2 represents solutions with light sleptons which can escape detection. Point 3 displays a solution that can remain viable even after LHC-Run3, despite improvements in the signal strength. Point 4 exemplifies a solution which potentially will be probed at the end of Run3 at $68\%$ CL. All these solutions reside in the slepton-neutralino coannihilation region in which the lighter chargino is mostly Wino-like.}
\label{tab2}
\end{table}

\section{Conclusions}
\label{sec:conc}
We revisit supersymmetric GUT models which impose non-universality in the SSB masses of the gauginos and matter fields at $\mgut$. Non-universality between the sfermion families removes the tension otherwise encountered between the Higgs boson mass and the muon $g-2$ anomaly, The third family is allowed to be heavy enough to realize a 125 GeV Higgs boson. Similarly, with non-universal gaugino masses imposed at $\mgut$, the impact on the muon $g-2$ contributions from the heavy gluino mass bound also disappears. However, the requirements for heavy squarks in the spectrum can still provide a tension in the muon $g-2$ region, since the first-two family sleptons and squarks are set to have the same mass ($m_{0_{1,2}}$) at $\mgut$. This tension can be ameliorated through the renormalization effects, and we present solutions for this class of supersymmetric GUT models that yield relatively light sleptons. The squarks and gluino can be as light as 3-4 TeV, and potentially accessible in Run3 experiments. 

We identify different sources for the SUSY contributions to muon $g-2$, in which the Bino usually plays a dominant role. In addition, the Wino also enhances the muon $g-2$ results to help reconcile the theory and the latest $g-2$ measurement by Fermilab. Although it is usually negligible, there is also a region of the parameter space in which the Higgsino contribution can play a significant role. The region providing a resolution to the muon $g-2$ anomaly within the $1\sigma$ interval bounds the first two family slepton masses at about 600-800 GeV, while the stau can be as heavy as about 3 TeV. These relatively low mass solutions are exciting since the LHC-Run3 should have the required sensitivity to probe them through the chargino-neutralino production processes. Such signal processes can provide a direct bound on the chargino mass, and we find that the muon $g-2$ solutions with $m_{\tilde{\chi}_{1}^{\pm}} \gtrsim 600$ GeV can be probed up to $68\%$ CL, and those with  $m_{\tilde{\chi}_{1}^{\pm}} \gtrsim 500$ GeV can be tested up to about $95\%$ CL. Considering the targeted luminosity (400 fb$^{-1}$) at the end of Run3, the charginos can be probed in Run3 up to about 700 GeV at $68\%$ CL and 600 GeV at $95\%$ CL. Even though these process do not provide a direct probe on the slepton masses, a larger portion of the region favored by muon $g-2$ will be probed at the end of the Run3 experiments. This provides a considerable impact when the LSP neutralino is Bino-like and the chargino is Wino-like. These solutions typically predict a relatively low spin-independent cross-section for dark matter scattering, and therefore further upgrades in the direct detection experiments are needed to test them. However, it is possible to realize solutions which yield relatively larger scattering cross-section ($\sim 10^{-10}$ pb). The relic density of LSP neutralino consistent with the Planck bound within $5\sigma$ can be realized through coannihilation processes consisting of the LSP neutralino together with the first two family sleptons, stau, and chargino. Among these solutions, the coannihilations involving sleptons and staus yield a relatively low scattering cross-section, but the chargino-neutralino coannihilation solutions predict an adequately large cross-section that can soon be tested.  In addition to the coannihilation scenarios, we also identify solutions in which the Planck dark matter bound can be satisfied through the $A-$resonance solutions.

\section*{Acknowledgments}
This work is supported in part by the United States Department of Energy grant DE-SC0013880 (QS) and by the Scientific and Technological research Council of Turkey (TUBITAK) Grant no. MFAG-118F090 (CSU). We acknowledge Information Technologies (IT) resources at the University Of Delaware, specifically the high performance computing resources for the calculation of results presented in this paper.

.

\bibliographystyle{JHEP}
\bibliography{mybibtex}

\end{document}